\documentclass[%
 aip,
 pop,
 amsmath,amssymb,
 reprint,%
citeautoscript
]{revtex4-2}

\usepackage{graphicx}
\usepackage{dcolumn}

\usepackage[utf8]{inputenc}
\usepackage[T1]{fontenc}

\makeatletter
\def\@email#1#2{%
 \endgroup
 \patchcmd{\titleblock@produce}
  {\frontmatter@RRAPformat}
  {\frontmatter@RRAPformat{\produce@RRAP{*#1\href{mailto:#2}{#2}}}\frontmatter@RRAPformat}
  {}{}
}%
\makeatother

\usepackage{array} 
\usepackage{booktabs}

\newcolumntype{L}{>{$}l<{$}}
\newcolumntype{C}{>{$}c<{$}}
\newcolumntype{d}[1]{D{.}{.}{#1}}

\usepackage{units}
\usepackage{subcaption}
\usepackage{stackengine}

\usepackage[dvipsnames, svgnames]{xcolor}
\colorlet{myIntColor}{DarkRed} 
\colorlet{myCiteColor}{DarkGreen} 
\colorlet{myExtColor}{blue}
\usepackage{hyperref} 
\hypersetup{colorlinks,
linkcolor=myIntColor,
citecolor=myCiteColor,
urlcolor =myExtColor,
menucolor=myIntColor,
}

\usepackage[capitalise,nameinlink]{cleveref} 

\usepackage[english]{babel}
\usepackage{csquotes} 
\usepackage{soul}
\graphicspath{{./}{figures/}}

\newcommand{\pD}[2]{\frac{\partial #2}{\partial #1}}

\newcommand{\D}[2]{\frac{{\rm d} #2}{{\rm d} #1}}

\newcommand\bb[1]{\mbox{\boldmath{$#1$}}}
\newcommand\grad{\bb{\nabla}}
\newcommand\bcdot{\bb{\cdot}}
\newcommand\btimes{\bb{\times}}

\newcommand{\mrm}[1]{\mathrm{#1}}
\newcommand{\ms}[1]{\mrm{#1}} 

\newcommand{\rmd}{{\rm d}}

\newcommand{\e}[1]{\ensuremath{\hat{\bb{#1}}}}
\newcommand{\const}{{\rm Const.}}

\newcommand{\Fh}{\ensuremath{F_\ms{h}}}
\newcommand{\Ft}{\ensuremath{F_\ms{t}}}
\newcommand{\Fs}{\ensuremath{F_\ms{s}}}

\newcommand{\ns}{\ensuremath{n_\ms{s}}}
\newcommand{\ncr}[1][]{\ensuremath{n_{\ms{cr}}^\ms{#1}}}

\newcommand{\qcr}[1][]{\ensuremath{q_{\ms{cr}#1}}}
\newcommand{\qa}{\ensuremath{q_a}}

\newcommand{\xf}{\ensuremath{x_\ms{f}}}
\newcommand{\It}{\ensuremath{I_\ms{T}}}
\newcommand{\Bs}{\ensuremath{B_\ms{s}}}
\newcommand{\Bg}[1][]{\ensuremath{B_\ms{g#1}}}
\newcommand{\Bpi}{\ensuremath{B_\ms{Pi}}}

\newcommand{\BPa}{\ensuremath{B_{\ms{P}a}}}

\newcommand{\Bti}[1][]{\ensuremath{B_\ms{Ti#1}}}
\newcommand{\BT}[1][]{\ensuremath{B_\ms{T#1}}}
\newcommand{\JT}{\ensuremath{J_\ms{T}}}
\newcommand{\JP}{\ensuremath{J_\ms{P}}}

\newcommand{\zapex}{\ensuremath{z_\ms{ap}}}

\newcommand{\hT}{\ensuremath{h_T}}

\newcommand{\RI}[1]{\ensuremath{I_\ms{#1}}}
\newcommand{\PCBenv}[1]{\ensuremath{\langle\delta \RI{#1}\rangle/I_\ms{p,max}}}
\newcommand{\zapInstab}{\ensuremath{\langle\delta z\rangle/\xf}}

\newcommand{\fluxT}[1]{\ensuremath{\Phi_\ms{#1}}} 

\newcommand{\ztrig}{\ensuremath{z_\ms{trig}}}
\newcommand{\ttrig}{\ensuremath{t_\ms{trig}}}

\newcommand{\NatureReproduce}{Reproduced from Myers \textit{et al.}, Nature \textbf{528}, 526 (2015). Copyright 2015 Nature Publishing Group.}
\newcommand{\PoPreproduce}{Reproduced from Myers \textit{et al.}, Physics of Plasmas \textbf{23}, 112102 (2016), with the permission of AIP Publishing}
\newcommand{\adaptedNote}{Chen, Astrophys. J. \textbf{338}, 453 (1989) and Chen and Krall, J. Geophys. Res.
Space \textbf{108}, 1410 (2003). Copyright 1989 American Astronomical Society
and 2003 American Geophysical Union.}

\renewcommand{\citet}[1]{ \citeauthor{#1} (\citeyear{#1})}
\newcommand{\citec}[1]{\cite{#1}} 
\newcommand{\PPPL}{Princeton Plasma Physics Laboratory, PO Box 451, Princeton, NJ 08543, USA}
\newcommand{\PU}{Department of Astrophysical Sciences, Princeton University, Peyton Hall, Princeton, NJ 08544, USA}

\definecolor{color1}{rgb}{0.0000,0.4470,0.7410}
\definecolor{color2}{rgb}{0.8500,0.3250,0.0980}
\definecolor{color3}{rgb}{0.9290,0.6940,0.1250}
\definecolor{color4}{rgb}{0.4940,0.1840,0.5560}
\definecolor{color5}{rgb}{0.4660,0.6740,0.1880}
\definecolor{redJet}{rgb}{.5,0,0}
\definecolor{blueJet}{rgb}{0,0,.5313}

\begin{document}

\title[Failed torus]{Laboratory study of the failed torus mechanism in arched, line-tied, magnetic flux ropes}







\author{Andrew Alt}
\email{aalt@pppl.gov}
\affiliation{\PU}\affiliation{\PPPL}
\author{Hantao Ji}
\affiliation{\PU}\affiliation{\PPPL}
\author{Jongsoo Yoo}
\affiliation{\PPPL}
\author{Sayak Bose}
\affiliation{\PPPL}
\author{Aaron Goodman}
\affiliation{\PU}\affiliation{\PPPL}
\author{Masaaki Yamada}
\affiliation{\PPPL}

\date{\today}

\begin{abstract}
    Coronal mass ejections (CMEs) are some of the most energetic and violent events in our solar system. The prediction and understanding of CMEs is of particular importance due to the impact that they can have on Earth-based satellite systems, and in extreme cases, ground-based electronics. CMEs often occur when long-lived magnetic flux ropes (MFRs) anchored to the solar surface destabilize and erupt away from the Sun. One potential cause for these eruptions is an ideal magnetohydrodynamic (MHD) instability such as the kink or torus instability. Previous experiments on the Magnetic Reconnection eXperiment (MRX) revealed a class of MFRs that were torus-unstable but kink-stable, which failed to erupt. These ``failed-tori'' went through a process similar to Taylor relaxation where the toroidal current was redistributed before the eruption ultimately failed. We have investigated this behavior through additional diagnostics that measure the current distribution at the foot points and the energy distribution before and after an event. These measurements indicate that ideal MHD effects are sufficient to explain the energy distribution changes during failed torus events. This excludes Taylor relaxation as a possible mechanism of current redistribution during an event. A new model that only requires non-ideal effects in a thin layer above the electrodes is presented to explain the observed phenomena. %
This work broadens our understanding of the stability of MFRs and the mechanism behind the failed torus through the improved prediction of the torus instability and through new diagnostics to measure the energy inventory and current profile at the foot points. 
\end{abstract}

\maketitle

Submitted to Physics of Plasmas

\section{Introduction}

The study of astrophysical phenomena is often accomplished via remote ground- and space-based observations and numerical simulations \citec{forbes2000,chen2017,green2018}. Many insights can be learned from both of these approaches but they also have their limitations and difficulties. The remote nature of observations limits the available diagnostics, and since one does not have control over the events, individual parameters cannot be adjusted to study the results in isolation. While simulations can control individual parameters and isolate specific phenomena, computational considerations require approximations that can limit the available physics. In this paper, laboratory experiments are used to create models of astrophysical phenomena so that they can be studied \textit{in-situ}. This allows for fine control of experimental parameters and for the use of a wide array of plasma diagnostics. The experiments presented here focus on the study of solar eruptions and the resulting coronal mass ejections (CMEs). These events lead to space weather which can be hazardous to Earth-based satellites as well as sensitive ground-based equipment \citec{pulkkinen2007}. Therefore, understanding and predicting space weather phenomena has been identified as a subject of great scientific interest \citec{board2013solar}. The experiments presented here allow for new physics insights that are required for better predictions of space weather events but are difficult to obtain otherwise.

CMEs often occur when long-lasting magnetic flux ropes (MFRs) protruding from the Sun suddenly and violently erupt \citec{crooker1997,green2009}. These arched structures are bundles of twisted magnetic-field lines anchored to the solar surface via line-tying to the conductive photosphere \citec{kuperus1974,chen1989,rust2003}. The trigger for an eruption in an MFR is often cast in terms of two ideal magnetohydrodynamic (MHD) instabilities, the kink and torus instabilities. These instabilities have long been studied in asymmetric devices such as tokamaks \citec{kruskal1954,bateman1978}, however they can also be extended to arched MFRs \citec{torok2004,kliem2006}. Previous experiments found a class of MFRs that were torus-unstable and kink-stable which failed to erupt \cite{myers2015}. These ``failed torus'' ropes would begin to rise, then the toroidal current would reorganize to become more hollow before the rope would collapse down, failing to erupt. Understanding the cause of the failed torus would help in future predictions of confined eruptions on the Sun. The previous description of the failed torus was that a self-organization event, such as Taylor relaxation, would occur \citec{taylor1974}. This would exchange toroidal and poloidal fluxes while maintaining the helicity, requiring non-ideal effects of magnetic reconnection.

The remainder of this paper has the following organizational structure: \Cref{sec:instabilities} describes the relevant forces and MHD instabilities that can occur in MFRs and then summarizes the previously observed failed torus regime. \Cref{sec:experiment} describes the experimental setup including the new diagnostics added for this work. Here, we will also present an overview of the parameter space that is reachable in our experiment. The remaining sections examine the Taylor relaxation explanation of the failed torus and find that based on the energy inventory (\cref{sec:LP-data}), that ideal MHD must hold during events. This rules out the possibility of Taylor relaxation which requires reconnection within the rope. \Cref{sec:Rogowski-data} validates the current hollowing throughout a failed torus rope and compares the foot-point current distribution of different regimes. \Cref{sec:measured-T-flux-change} then presents a mechanism through which we can explain the changes in toroidal flux without violating ideal MHD except for a small region above the foot points. \Cref{sec:flux-indexed-values} examines the change in toroidal field profile during events. Together, these provide a sufficient mechanism to explain the failed torus. This section also discusses how these results can be extended to the conditions on the Sun.

\section{Flux ropes and associated instabilities}
\label{sec:instabilities}

\begin{figure}
    \centering
    \includegraphics[width=\linewidth]{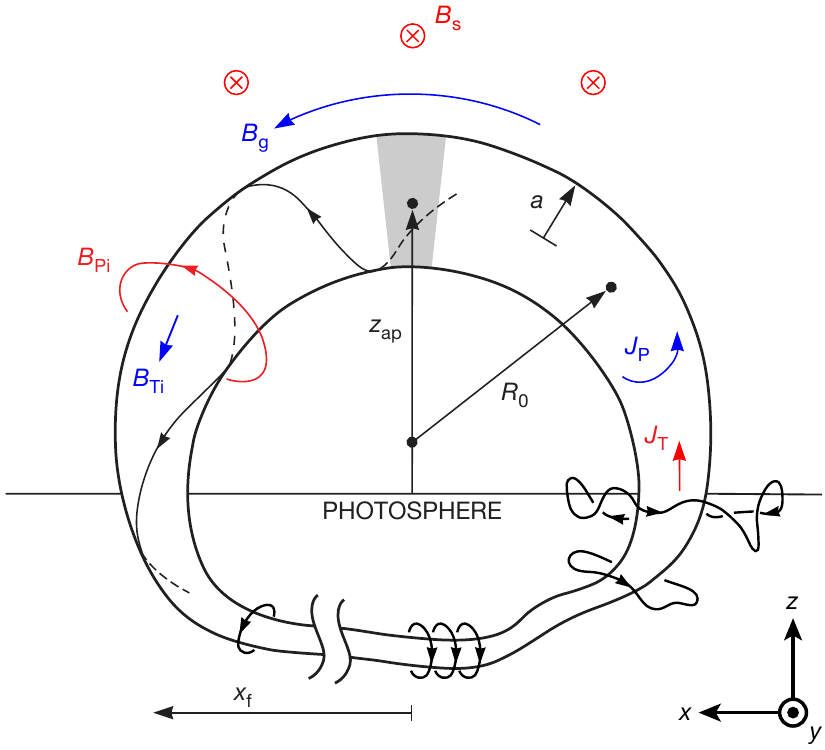}
	\caption{A model of an arched, line-tied magnetic flux rope, showing the breakdown of the fields and currents based on the direction and source. The line-tying condition at the conducting photosphere causes the foot points to be anchored with a separation of $2\xf$. The internal fields, \Bpi{} and \Bti, are generated by the currents in the rope, while the external fields, \Bs{} and \Bg, are generated by the Sun. \PoPreproduce, and adapted with permission from \adaptedNote }%
    \label{fig:chen-model}
\end{figure}

\renewcommand{\arraystretch}{1.5}
\begin{table*}[!htbp]
	\centering
\begin{tabular}{ @{}lccc@{} }\toprule 
	Force  &  Symbol     & Source Term & Analytical Expression \\
	\midrule
	Hoop force (upward) & \Fh     & $f_\ms{h}=\JT\Bpi$             & $\Fh=\frac{\mu_0\It^2}{4\pi R}\left[\ln\left(\frac{8R}{a}\right)-1+\frac{\ell_\ms{i}}{2}\right]$ \\ 
	Strapping force (downward) &\Fs & $f_\ms{s}=-\JT\Bs$             & $\Fs=-\It B_\ms{s0}$ \\ 
	Tension force (downward) & \Ft  & $f_\ms{t}=-\JP\BT$ 	& $\Ft=\frac{-\mu_0\It^2}{8\pi R}\left[\frac{\left< B_\ms{T}^2\right>-B_\ms{g0}^2}{B_{\ms{P}a}^2} \right] \approx -\frac{1}{2}\frac{\mu_0\It^2}{4\pi R} $\\ 
	\bottomrule
\end{tabular}
\caption{Breakdown of forces on a flux rope. The forces are separated by contributions to the $\bb{J}\btimes\bb{B}$ term. The fields and currents used are shown in \cref{fig:chen-model} and the analytical expressions are derived in \citet{myers2016}. \PoPreproduce%
}
\label{tab:forces}
\end{table*}

In an MFR, the magnetic field, \bb{B}, and electric currents, \bb{J}, of an MFR can be broken up into components based on their direction and source as shown in \cref{fig:chen-model}. First, \bb{J} can be divided into toroidal, T, and poloidal, P, components. Then, \bb{B} can be separated into two components: internal and external. The internal field is generated by currents within the MFR, while the external field is generated by currents within the Sun. The internal field is further divided into toroidal and poloidal components. The external field is divided into the guide field, \Bg, along the axis of the rope and the strapping field, \Bs, perpendicular to this axis. 

The dominant force on a low-$\beta$ plasma, such as an MFR, is the $\bb{J}\btimes\bb{B}$ force. Here, $\beta\equiv2\mu_0 P/B^2$ is the ratio of thermal to magnetic energy in the plasma, $P$ is the plasma pressure, and $\mu_0$ is the permeability of free space. This force can be decomposed based on the components of the source terms discussed above. These forces are called the hoop, strapping, and tension forces and are defined in \cref{tab:forces}. The tension and strapping forces hold a rope down while the hoop force causes it to expand upward.

\subsection{The tension force}

\label{sec:tension-force}
Of particular importance to this work is the tension force, \Ft, caused by the poloidal current and toroidal field, $f_\ms{t}=-\JP\BT$. The tension force is caused by the helical winding of current around a torus. The force in the direction of the major radius, $R$, is directed inward on the inboard side of the torus and is directed outward on the outboard side. However, since the number of current windings must be the same on both sides, both \JP{} and \BT{} are larger on the inboard side. When the $R$-directed force is averaged over a poloidal cross section, the net force is directed inward to the center of the torus.

When \Ft{} is simplified using the large-aspect-ratio approximation in a full torus, it is given by \citec{alt2022thesis}
\begin{equation}\label{eqn:FTpara}
    \Ft = -\frac{\mu_0 \It^2}{8\pi R_0}\left[\frac{\left\langle\BT^2\right\rangle - \Bg[0]^2}{\BPa^2}  \right] , 
\end{equation}
where $\langle\dots\rangle$ denotes cross-section averaging, $R_0$ is the major radius, \It{} is the total toroidal current, \Bg[0] is the edge guide field, \mbox{$\BPa\equiv\mu_0\It/(2\pi a$)} is the edge poloidal field, and $a$ is the minor radius. Here we see that a downward tension force is caused by a paramagnetic toroidal field within the rope.

\subsection{The torus instability} \label{sec:TI}

The torus instability occurs when the net force on an MFR increases as it is displaced upward away from equilibrium, \citec{bateman1978,forbes1991,kliem2006}
\begin{equation}\label{eqn:TI_force_balance}
    \left.\sum_i F_i=0\right|_{z=\zapex} \quad \text{and} \quad \left.\sum_i\pD{z}{F_i}>0\right|_{z=\zapex} ,
\end{equation}
where $F_i$ are the constituent forces given in \cref{tab:forces} and \zapex{} is the equilibrium height of the rope's apex. It is sometimes described as a loss-of-equilibrium via a ``catastrophe'' mechanism \citec{forbes1991}.

The torus instability was first studied in the context of fusion devices, where the role of the external strapping field, \Bs, is played by the applied vertical field \citec{bateman1978}. Since the main inward force is caused by \Bs{}, the torus instability criterion is often cast in terms of the decay index of this field,\footnote{The decay index is defined such that a field, $B\propto z^{-n}$ has a decay index of $n$.}
\begin{equation}\label{eqn:ns_def}
\ns=-\frac{z}{\Bs}\pD{z}{\Bs} > \ncr,
\end{equation}
where $z$ is the height above the photosphere and $\ncr=1.5$ in an axisymmetric, large aspect ratio, full torus \citec{kliem2006}. When previous MFR experiments were translated to the conditions on the Sun, a lower value of $\ncr\approx0.9$ has been observed in the partial tori of arched MFRs \citec{alt2021}. If \Bs{} decays faster than this critical value, the torus instability can occur.

\begin{figure*}
    \centering
    \includegraphics[width=\linewidth]{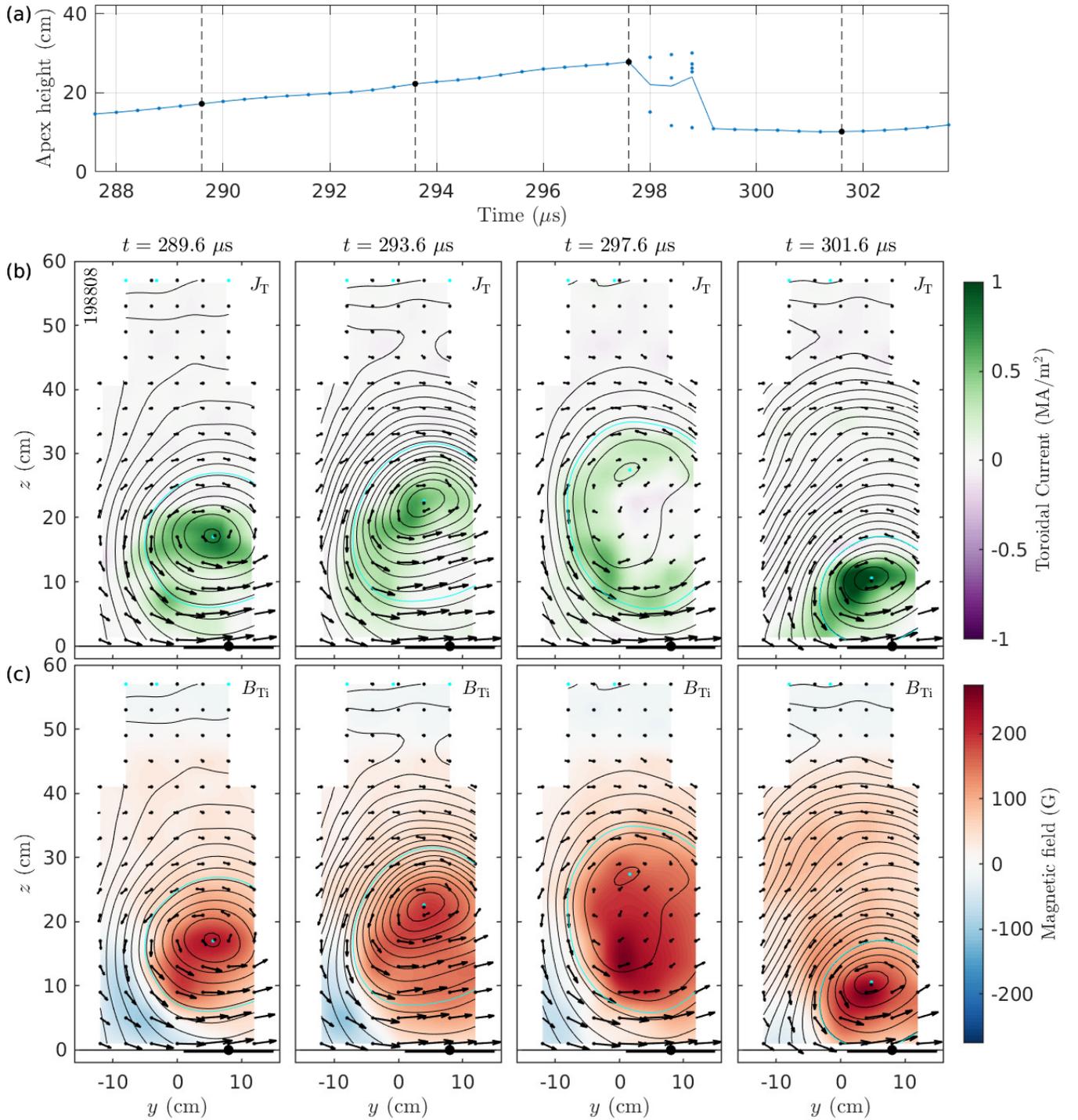}
    \caption{Example data taken during a failed torus event. \textbf{a}: The apex height during the failed torus event. The points represent heights of the nulls of the poloidal magnetic field at each time. \textbf{b}, \textbf{c}: 2D plots of the toroidal current density (top) and toroidal magnetic field (bottom) at the times indicated in \textbf{a}. The in-plane magnetic field at the probe locations is also shown via arrows.}
    \label{fig:FT_time_slice}
\end{figure*}

\subsection{The kink instability} \label{sec:KI}

The kink instability is another ideal MHD instability that can affect MFRs \citec{kruskal1954,shafranov1956,torok2004}. The instability occurs when \It{} is too large, or equivalently, \Bg{} is too small, such that the field lines at the edge of the rope become too twisted around the axis. When the field lines at the edge of the rope become too twisted around the axis, the instability can occur. This twisting can be described in terms of the edge  safety factor, $\qa$, or equivalently, the twist number $N$, 
\begin{equation}
    \qa \equiv \frac{1}{|N|} = \frac{2\pi a}{L}\frac{\Bg[0]}{\BPa}  < \qcr ,
\end{equation}
where $L$ is the length of the MFR.
For toroidally symmetric fusion devices, the critical safety factor is $\qcr=1$ \citec{kruskal1954,shafranov1956}. However in previous MFR experiments where toroidal symmetry is broken by line-tying, a lower critical value of $\qcr\approx0.8$ has been observed \citec{myers2016}. When the kink instability onsets, it causes the axis of a rope to tilt and begin to rotate. However, the instability saturates nonlinearly at relatively small amplitudes and therefore cannot be the sole cause of eruptions.

\subsection{Observations of failed tori in previous experiments}
\label{sec:FT-description}

In a previous experimental campaign, a class of ropes that were torus-unstable but kink-stable were observed failing to erupt \citec{myers2015}. This regime is referred to as the ``failed torus'' regime and an example MFR is shown in \cref{fig:FT_time_slice}. Since the rope is initially unstable to the torus instability, the apex begins to rise. However, before a full eruption can occur, the toroidal current profile becomes more hollow, causing the toroidal flux to increase. This flux increases causes a spike in the tension force (due to \cref{eqn:FTpara}), and the rope collapses back down to a lower height \citec{myers2017PPCF,myers2015}.

The reorganization that occurs and increases the toroidal flux simultaneously decreases the poloidal flux. This can be explained by the conservation of helicity which describes the linking of these two fluxes. Magnetic helicity measures the linking of magnetic-field lines and is given by \citec{taylor1974,berger1984} $H=\int\bb{A}\bcdot\bb{B} \rmd V$, where \bb{A} is the vector potential defined such that $\grad\btimes\bb{A}=\bb{B}$. In order for $H$ to be gauge invariant, the integral must be carried out over a simply connected volume, $V$, whose boundary, $S$, is a magnetic surface such that $\left.\bb{B}\bcdot\e{n}\right|_S=0$, where \e{n} is the unit normal vector to $S$. Helicity conservation is a common feature of Taylor relaxation \citec{taylor1974,taylor1986} where the global field-line linking is conserved while the total energy is minimized. Under this constraint, the minimum energy state is a force-free field where $\mu_0\bb{J}=\alpha\bb{B}$, with $\alpha$ being a constant. However, in order to reach this state, field-line breaking via reconnection must be allowed in the plasma volume.

\section{Experimental setup}
\label{sec:experiment}

\begin{figure}%
    \centering
    \includegraphics[width=\linewidth]{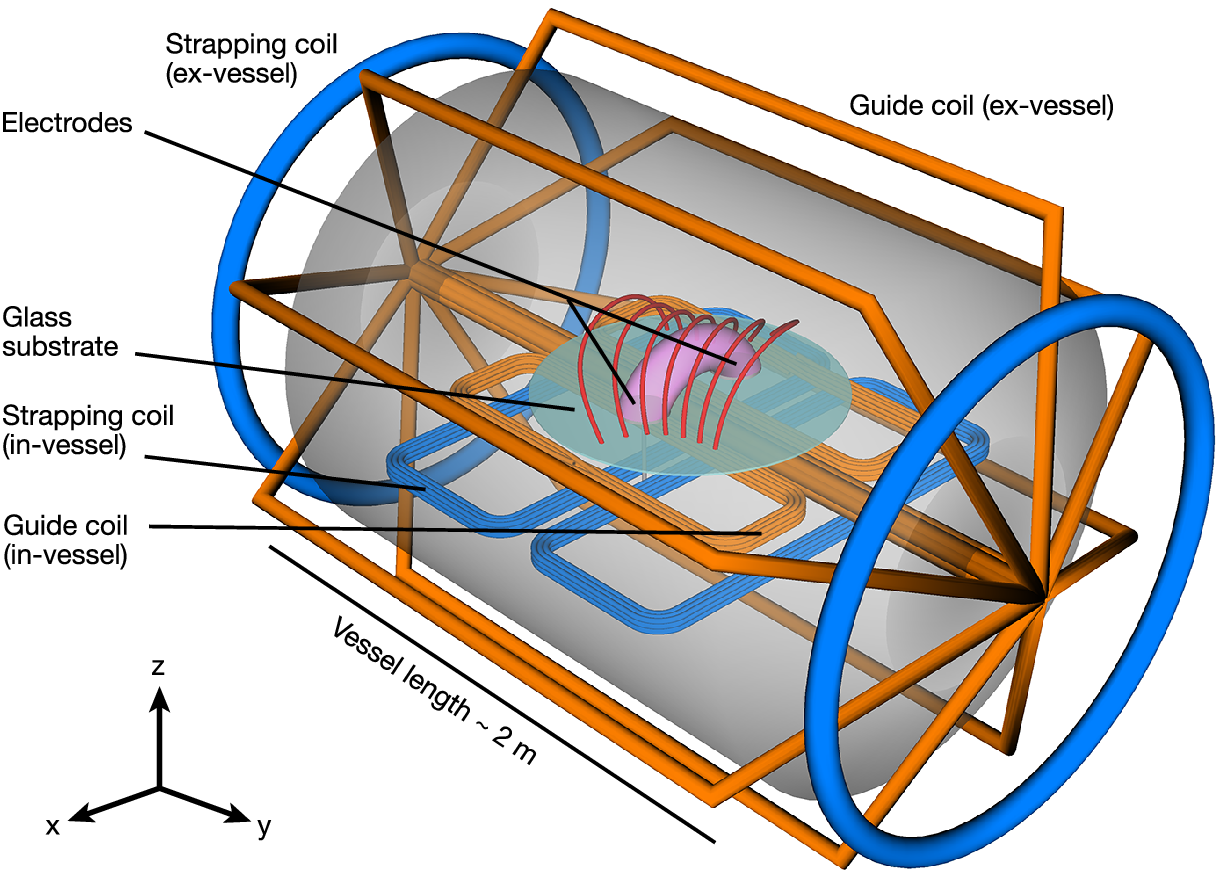}
	\caption{The MRX vessel used to create arched, line-tied flux ropes. An arc discharge is created between two copper electrodes and is separated from the magnetic-field coils by a glass substrate. The model in \cref{fig:chen-model} corresponds to the pink plasma arc in the center of the image. Four coils were inserted into MRX in order to control the profiles of both the guide and strapping fields. The orange coils contribute to the guide field along the rope while the blue coils control the strapping field across it. Control of the vacuum fields allows for control of the instability parameters for the torus and kink instabilities. \NatureReproduce}%
    \label{fig:MRX}
\end{figure}

Flux ropes have previously been created inside Magnetic Reconnection Experiment (MRX) in order to study their solar counterparts \citec{yamada1997,myers2015,myers2016,alt2021,alt2022thesis}. The experimental setup is shown in \cref{fig:MRX} and is described in further detail in \citet{myers2016}. The vacuum fields are controlled by four sets of coils, two for each the guide and strapping fields. By controlling the magnitude and direction of the current in these coils, the strength and decay index of these two fields can be controlled independently. The currents evolve slowly compared to a rope's lifetime, $\tau_\ms{MFR}\sim\unit[1]{ms}$. The ropes are created by a discharge between two copper electrodes with radius $a_0=\unit[7]{cm}$ with a foot-point half-separation of $\xf=\unit[17]{cm}$. The electrodes form the foot points of a rope and can be considered to be perfectly conducting on the timescale of a rope's lifetime. Therefore the ropes are line-tied to their foot points and exist within a static background magnetic field. Before the discharge is initiated, the vessel is filled with neutral hydrogen to a pressure in the range of $\unit[30-40]{mTorr}$.

An important property of solar MFRs is the large separation between both time and spatial scales. For the storage and release paradigm of CMEs to hold, the characteristic driving time for energy buildup in the corona, $\tau_\ms{D}$, must be much larger than the resistive dissipation time, $\tau_\ms{R}\equiv\mu_0 aL/\eta$. In addition, for MHD to be valid, the Alfv\`en transit time, $\tau_\ms{A}\equiv L/v_\ms{A}$, must be much shorter than both of these timescales, i.e. $\tau_\ms{R}\gg\tau_\ms{D}\gg\tau_\ms{A}$. In our experiments, these times are $\tau_\ms{R}\sim\unit[1]{ms}$, $\tau_\ms{D}\sim\unit[150]{\mu s}$, and $\tau_\ms{A}\sim\unit[3-10]{\mu s}$, and therefore the inequality holds \citec{alt2022thesis}.

The primary diagnostic in the flux rope experiments is an array of B-dot probes with over 300 pickup coils arranged in a 2D array (usually placed in the $y$-$z$-plane) with two lines of out-of-plane probe triplets along the center line. The out-of-plane probes are placed parallel to the $z$-axis at $y=0$ and $x=\unit[\pm 3]{cm}$. The out-of-plane probes give better insight into the instantaneous curvature than was previously possible. The curvature measurements yield an estimate for the out-of-plane derivatives of \bb{B} which are necessary to calculate the in-plane components of \bb{J}.

\subsection{Rogowski coils for measuring the foot-point current distribution}
\label{sec:Rogowski-description}

\begin{figure}
    \centering
    \begin{subfigure}{.55\linewidth}
        \includegraphics[width=\linewidth]{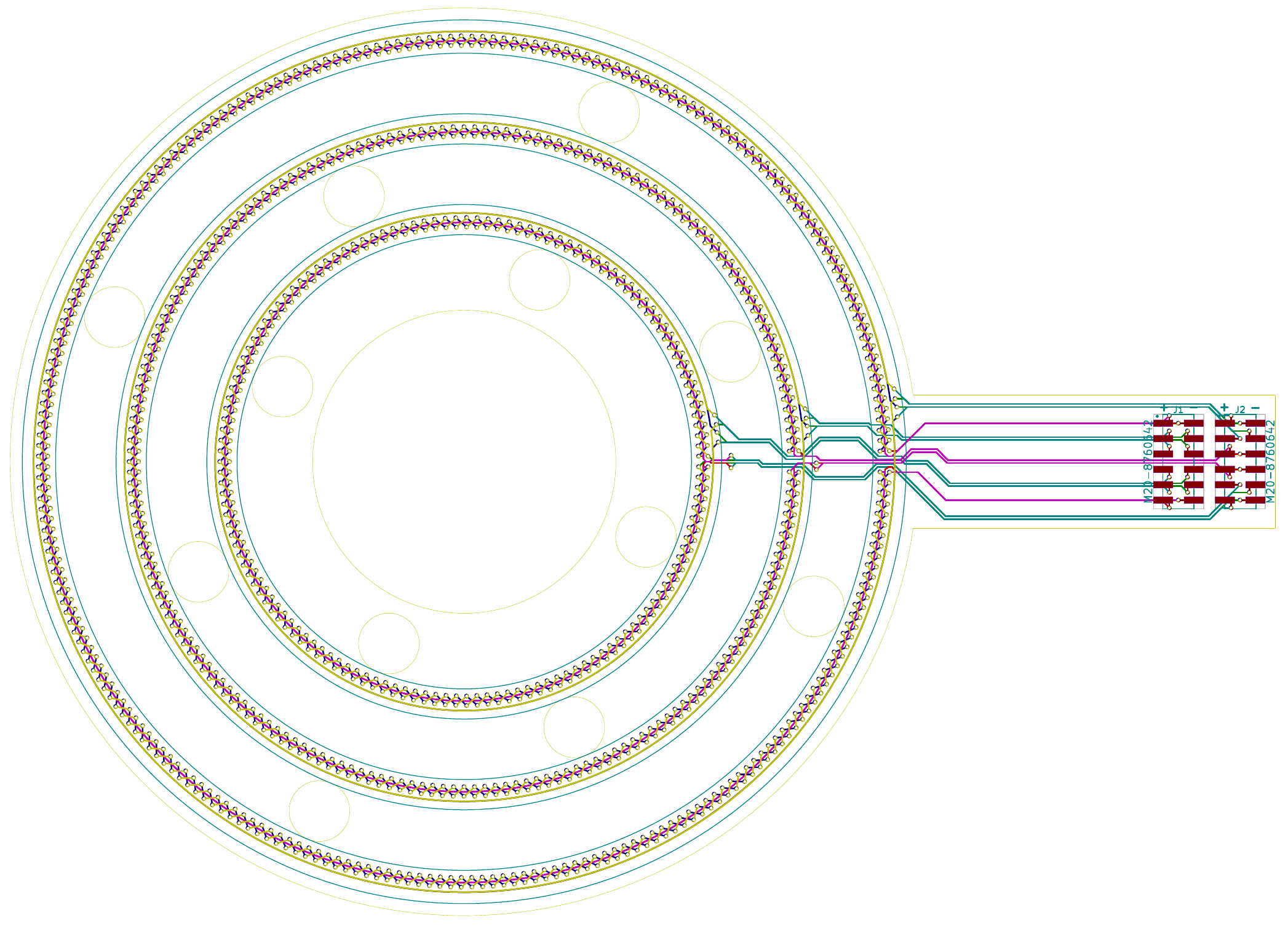}
        \phantomsubcaption\label{fig:PCB_kicad}
    \end{subfigure}%
    \hfill
     \begin{subfigure}{.44\linewidth}
        \includegraphics[width=\linewidth]{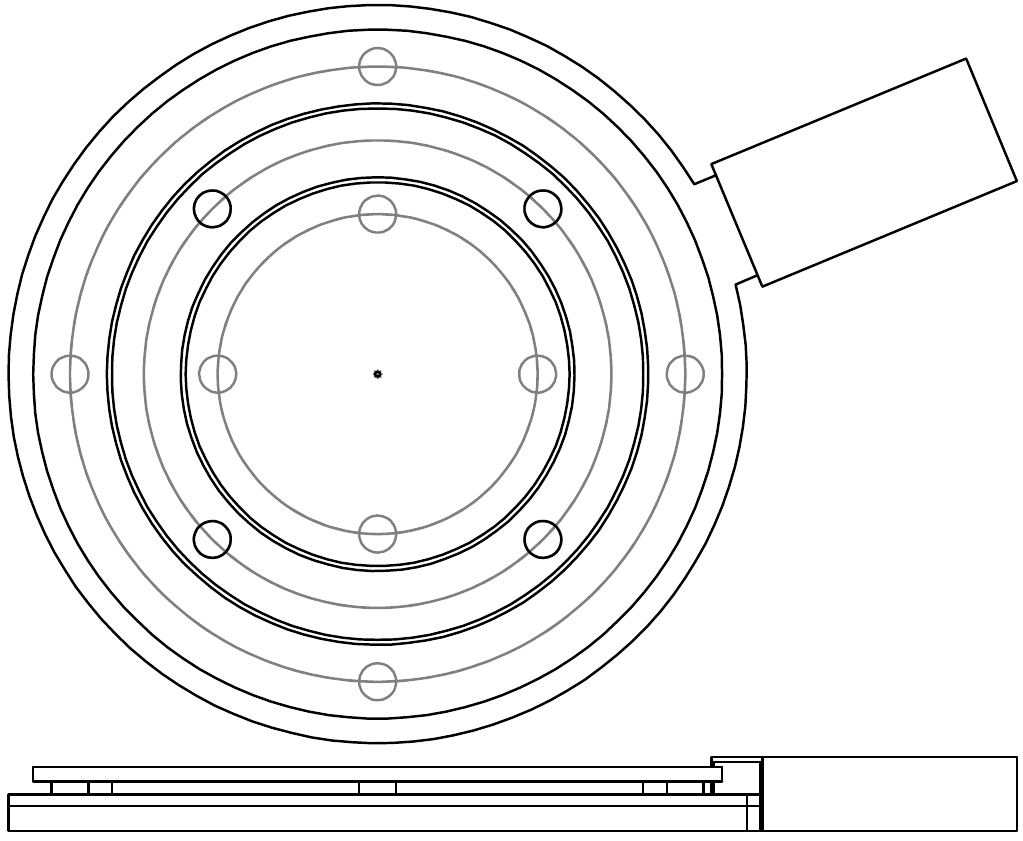}
        \phantomsubcaption\label{fig:PCB_cad}
    \end{subfigure}%
    \caption{Diagram of the Rogowski coils used to measure the distribution of plasma current at the foot points. \textbf{Left}: The circuit that was printed into a PCB to create the three concentric Rogowski coils. \textbf{Right}: Top and side view of the positioning of the Rogowski coils below the segmented copper electrodes. The ``panhandle'' contains the electrical connections used to measure the signals which were insulated from the plasma. A small hole was made in the center of the electrode to allow for gas puffing at the foot points.}\label{fig:PCB}
\end{figure}

Rogowski coils were made via a printed circuit board (PCB) (see \cref{fig:PCB}) and were placed below the electrodes located at the foot points of the MFRs \citec{kojovic2003}. There were three sets of coils printed into each board and arranged to measure the current passing through the three segments of the electrodes. Each electrode segment consists of a concentric circle or annulus of copper and four posts connecting each to a shared base plate which is connected directly to the capacitor bank driving the flux ropes. The PCBs were then placed between the copper segments of the electrode and the base plate. Each set of Rogowski coils consists of two counter-wound coils which is used to eliminate electrostatic noise. Additionally, flux loop pairs were printed into each PCB to measure the toroidal magnetic flux at each coil location. 

The coils were calibrated by placing them around a wire that was connected to the driving capacitor bank to create currents similar in magnitude and timescale to the experimental currents. The calibration current was externally measured with a commercially available Rogowski coil. Previous measurements of failed torus ropes were focused on the apex of the rope and could not measure any potential hollowing of the current at the foot points. The addition of segmented electrodes and Rogowski coils at the foot points allows for the measurement of the current distribution at the foot points to determine whether hollowing is occurring throughout the entire rope or if it is localized to the apex.

\subsection{Langmuir probes in the flux rope experiments}
\label{sec:LP-description}
Triple Langmuir probes have been used extensively in previous MRX operation to measure the plasma density, temperature, and potential \citec{yoo2013thesis}. However, in previous experience with the flux rope setup, it has been seen that the probe tips quickly became dirty and it was not feasible to remove and clean them often enough to maintain data quality. To combat this, a new type of probe tip was created. These new probe tips consist of a loop of 0.005" diameter tungsten wire that is held in place by a double-bore alumina tube. Each end of the loop is connected to a twisted pair that runs out of the vacuum vessel to the probe circuit. During measurement, each twisted pair is shorted together, allowing each loop to be biased as a uniform probe tip. Periodically, a current of about $\unit[2]{A}$ was passed through each loop until it began to faintly glow from black-body radiation. This heating removed surface impurities via outgassing, improving performance. The area of each probe tip was measured by cross-calibrating with a reference Langmuir probe in the regular MRX configuration with a quiet, toroidally symmetric plasma \citec{yoo2013thesis,alt2022thesis}.

\subsection{Experimental parameter space}\label{sec:param-space}

\begin{figure}
	\centering
	\includegraphics[width=\linewidth]{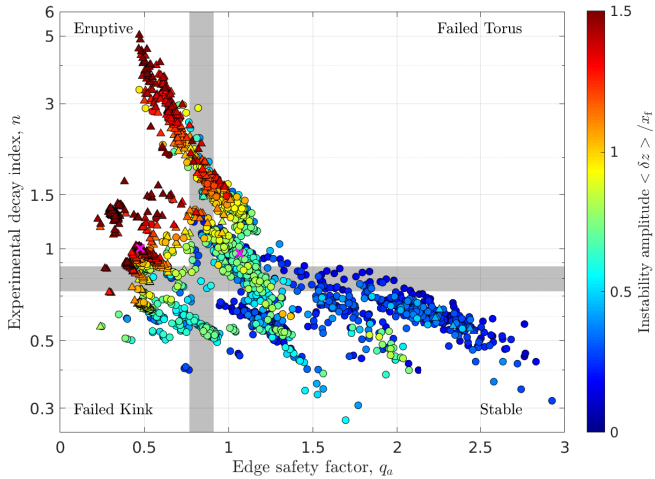}
	\caption{The experimental parameter space of nearly 2,500 shots. Each point represents a shot and is placed based on its value of the two ideal instability parameters, \qa{} and $n$. The color represents the instability amplitude based on the magnitude of repeated deviations of \zapex{}, normalized by \xf. Shots that experienced multiple eruptions are represented with triangles while non-eruptive shots are circles. The magenta crosses represent specific parameters that will be considered in \cref{sec:LP-data}. Four different regions of stability can be seen, though their boundaries are not perfectly defined by the parameters.}
    \label{fig:ns_vs_q}
\end{figure}

In \cref{fig:ns_vs_q}, each shot taken during three separate experimental campaigns is plotted based on its value of the stability parameters for the torus and kink instabilities, $n$ and \qa{} as described in \cref{sec:KI,sec:TI}. The results of nearly 2,500 discharges are represented in the plot. The color of each point represents the instability amplitude, \zapInstab, which is based on the time-averaged size of an envelope around the apex height changes normalized by the foot-point half-separation, \xf{}. (The instability amplitude is described in further detail in \citet{myers2016}.) Shots that erupted multiple times are shown by triangles while non-eruptive shots are circles.

\subsubsection{The quadrants of the experimental parameter space}
\label{sec:quadrant_description}

The four quadrants in \cref{fig:ns_vs_q} can be described by qualitatively different stability properties, as labeled. The boundaries of these quadrants are denoted by the gray bars centered on $\qa=0.8$ and $n=0.8$. Ropes in the ``Stable'' quadrant are stable to both the kink and torus instabilities and do not erupt. Ropes in the ``Failed Kink'' quadrant have a lower \qa{} and are unstable to the kink instability. However, since the kink instability saturates at a relatively low amplitude, these ropes kink and rotate without erupting. In the ``Eruptive'' quadrant, ropes have high $n$ and low \qa{} and therefore are unstable to both the torus and kink instabilities. These ropes are able to erupt. The last regime is the ``Failed Torus'' quadrant, which contains ropes that are torus-unstable but kink-stable. Despite being torus-unstable, these ropes often fail to erupt (see \cref{sec:FT-description} for more detail). 

The apex height (Defined as nulls of the poloidal magnetic field.) of an example eruptive rope is presented in \cref{fig:LP-time}. Here it can be seen that \zapex{} is multivalued at certain times. This occurs due to experimental constraints caused by the large external inductance in the circuit driving the plasma current. This external inductance causes the total toroidal current to remain constant on the timescale of an eruption. Therefore, instead of the erupting rope dissipating as it rises, a new rope is formed at a lower height. This rope exists in the same vacuum fields, and is therefore also unstable. The new rope then erupts, creating a repeating pattern of eruptions. At times when two ropes coexist, \zapex{} is presented as multivalued and usually consists of three values, two O-points corresponding to the two ropes' apexes and an X-point between them.

\section{Evolution of the energy inventory during events}
\label{sec:LP-data}

In this section we are primarily focused on the energy inventory before and after a failed torus event. Since our experimental plasma is low-$\beta$, the majority of the stored energy is in the magnetic fields, however the thermal and kinetic energies are significant enough to be considered. The thermal properties were not able to be measured on previous flux rope experiments conducted on MRX due to difficulties of implementing Langmuir probes in the flux rope environment.
The triple Langmuir probe design described in \cref{sec:LP-description} allowed for a direct measurement of the electron thermal properties.

\subsection{Event identification and timing}\label{sec:LP-time}

\begin{figure}
    \centering
    \includegraphics[width=\linewidth]{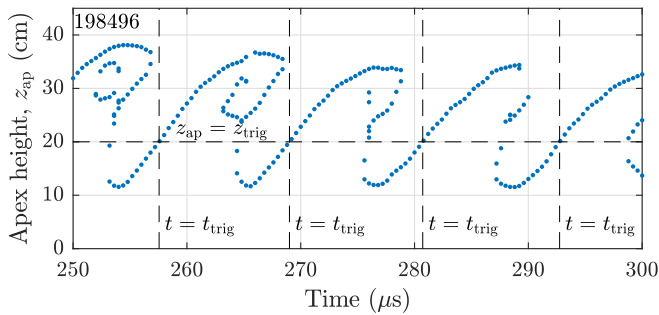}
    \caption{The data overlaying procedure for an example eruptive rope. The apex height, \zapex, at each time is shown by the blue points. Whenever the rope crosses a set trigger height, $\ztrig=\unit[20]{cm}$ (horizontal dashed line), the crossing time, \ttrig{} (vertical dashed lines), is noted. The data in a fixed window around each \ttrig{} is then averaged together to build the behavior of an ``average event.''}
    \label{fig:LP-time}
\end{figure}

In this section, we will be focused on what happens during an event, which will be defined as either an eruption or a failed torus event. In order to investigate the behavior of an ``average event,'' we look first at the data gathered from many shots with the same experimental conditions. However, due to the nature of the instabilities, even under the same experimental conditions, events can occur at very different times in the shot. While the events may occur at very different times, they are often very similar if correctly shifted in time. The procedure for averaging event data is shown with the values of \zapex{} presented in \cref{fig:LP-time}. First a fixed height, $\ztrig=\unit[20]{cm}$, is picked.\footnote{The specific value of \ztrig{} is not important to the final results as long as it falls within the range of most events.} Then each time, \ttrig, where the rope passes through this height is found, i.e. $\zapex(t=\ttrig)=\ztrig$. After each \ttrig{} is found, the value of $\zapex(t+\Delta t)$ is recorded for time shifts, $\Delta t\in [\unit[-20]{ms},\unit[20]{ms}]$. This data is then averaged for each $\Delta t$ across events and shots to create an average event. The averaging procedure can be repeated for any other experimentally measured quantity such as the results of the Langmuir probes.

The averaged apex height for a selected failed torus condition and an eruptive condition are shown in the bottom row of \cref{fig:LP-ave-Etot}. The times where there is no apex for the eruptive condition correspond to times where many of the events have multiple apexes and a good average cannot be created. The repetitive nature of the eruptive events can be seen in the events both preceding and following the event occurring at $\Delta t=0$. These events survive the averaging due to the coherent nature of the eruptive events across shots. In contrast to this, only one event can be seen in the failed torus condition. This is because the events are not as coherent and the delay between subsequent events varies significantly between shots, even with the same experimental conditions. However, if we limit our analysis to around $\unit[-5]{\mu s}<\Delta t<\unit[5]{\mu s}$, the events remain more coherent and averaging can be used to build an average event.

\subsection{Measurements of an average event}

We first want to investigate how the total energy stored within a flux rope changes during an event. The stored energy density is given by\footnote{Due largely to experimental constraints, we have chosen to neglect the ion pressure in this equation. The consequences of this are discussed in \cref{sec:LP-E-breakdown}.}
\begin{equation}
    w=\frac{P_e}{\gamma-1} + \frac{B^2}{2\mu_0} 
    ,
\end{equation}
where $P_e$ is the electron pressure and $\gamma$ is the adiabatic index and we will use $\gamma=5/3$. Therefore the total energy per unit length in a wedge area around the apex is given by
\begin{equation}
    W=\int\hT\left[\frac{P_e}{\gamma-1} + \frac{B^2}{2\mu_0} 
    \right]\rmd A,
\end{equation}
where the integral is carried out in the poloidal plane over the flux surface that contains 50\% of the current and \hT{} is the scale factor defining the width of the wedge volume around the apex (highlighted in \cref{fig:chen-model}) as described in \citet{alt2022thesis}. The flow energy of a moving flux rope can be defined as the kinetic energy of the ions moving with the flow velocity,
\begin{equation}
    K=\int\hT\left[\frac{1}{2}m_i n v_\ms{ap}^2 \right]\rmd A,
\end{equation}
where $m_i$ is the ion mass, $n$ is the particle density, and $v_\ms{ap}$ is the apex velocity.

\begin{figure}
    \centering
    \includegraphics[width=\linewidth]{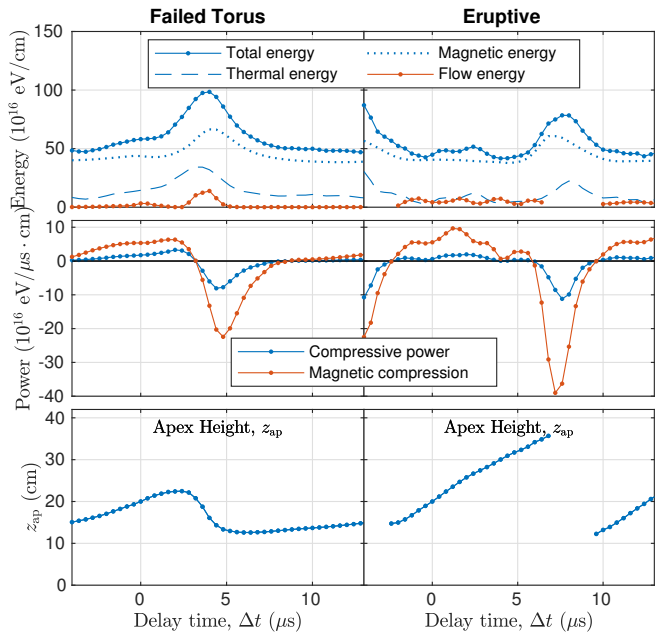}
    \caption{Energy inventory during an average failed torus (left) and eruptive (right) event. \textbf{Top:} The internal pressure energy is shown with a blue dashed line, the magnetic energy is a blue dotted line, and the sum of these is a solid blue line. The flow energy of the moving apex is shown in red. \textbf{Middle:} Compressive power done by the rope, defined by \cref{eqn:E_conservation}. The two terms of this equation are plotted separately. The compression due to the thermal pressure is shown in blue in the top row and the compression of the magnetic field is shown in red. \textbf{Bottom:} The average apex height.}
    \label{fig:LP-ave-Etot}
    \label{fig:LP-compressive}
\end{figure}

The average value of $W$ and the two terms separately are presented in the top row of \cref{fig:LP-ave-Etot} along with the flow energy for both a failed torus and eruptive condition. The pressure and density in the rope are measured via the triple Langmuir probe as described in \cref{sec:LP-description} and the magnetic energy is measured with the B-dot probe array. In the failed torus case, we see a large change in the total energy stored in the rope during the collapse, while there is a more moderate change in the eruptive case. At times where the eruptive ropes split and there is no well defined apex height, the flow energy cannot be defined.

The scale factor (and therefore wedge width) used in \cref{fig:LP-ave-Etot} is normalized by setting $\hT(\zapex)=1$ so that the integral is scaled by a unit length. In addition to this scaling, the changing width with height is compensated by setting the nominal wedge width to be a fixed fraction of the total rope length using the shifted circle model described in \citet{alt2021}. This fixed fraction compensates for changes in energy per unit length that would be caused by the changing of the total rope length.

In addition to the stored energy within a rope, we also want to investigate the work done on a rope as events occur. When a failed torus rope collapses, it moves into a region of larger external magnetic field, which causes the ropes to decrease in cross-sectional area. The change in volume of the wedge around the apex causes work to be done on the rope via \cref{eqn:E_conservation} as shown in \cref{sec:E-conservation}. One of the assumptions made in \cref{sec:E-conservation} is that the integration volume follows the plasma velocity. Since we are interested in a volume centered on the apex, this assumption is equivalent to neglecting axial flow. We are interested in Alfv\'enic timescales, therefore the flow is negligible as long as $v_\ms{flow}\ll v_\ms{A}$, which is expected in our experiments based on the frequency of the kink instability 
\citec{oz2011,ryutov2006}.  

The compressive power done by a rope is then given by 
\begin{equation}
    P_\ms{comp}= \left(P_e + \frac{B^2}{2\mu_0}\right)\D{t}{V},
\end{equation}
where $\rmd V/\rmd t$ is the rate of change of the wedge volume around the apex defined by \hT{}, described in \citet{alt2022thesis}. The measured compressive power is shown in the middle row of \cref{fig:LP-ave-Etot}. As a failed torus rope rises, it does a small amount of work in order to expand in cross-sectional area. However, when the rope quickly collapses during a failed torus event, a large amount of work is done on the rope. The same values of $B$ is used in the calculation of the magnetic terms of both $W$ and $P_\ms{comp}$.

Similar compressive power calculations can be done for average eruptive ropes. However, the decrease in apex size of an eruptive rope before an event is due to the creation of a new, lower rope (as discussed in \cref{sec:quadrant_description}) and not due to an actual decrease in size. Therefore, the change in apex size cannot be considered a true compression and the values plotted in \cref{fig:LP-compressive} for these times are not accurate.

\subsection{Energy inventory during a failed torus event} 
\label{sec:LP-E-breakdown}

\renewcommand{\arraystretch}{1.1}
\begin{table}
    \centering
    \begin{tabular}{l| c| d{2.1}}\toprule
         Work or energy change & Source term & \multicolumn{1}{c}{Energy ($\unit[10^{16}]{eV/cm}$)}  \\
         \midrule
         Pressure compressive & $\int P_e\D{t}{V}\rmd t$  & -8.1\\
         Magnetic compressive & $\int \frac{B^2}{2\mu_0}\D{t}{V}\rmd t$  & -57\\
         Thermal energy & $\int\frac{P_e}{\gamma-1}\rmd V$ & 12 \\
         Magnetic energy & $\int\frac{B^2}{2\mu_0}\rmd V$ & 36 \\
         Flow energy & $\int\frac{1}{2}\rho v^2\rmd V$ & 13 \\
         \midrule
         Total work & & -65 \\
         Total energy change &  &  62 \\
         \bottomrule
    \end{tabular}
    \caption{The energy inventory during an averaged failed torus event. The work done on the rope by the changing volume as well as the energy within a wedge around the apex are considered. The work is integrated over the same time window the difference in energies are taken, $\unit[3.2]{\mu s}\le\Delta t\le\unit[8.0]{\mu s}$. The work done and energy change balance within 6\%, implying that ideal MHD is sufficient to describe the energy change during a failed torus event.}
    \label{tab:LP_E_breakdown}
\end{table}
\renewcommand{\arraystretch}{1.0}

The energy inventory during an average failed torus event can be found via the plots in \cref{fig:LP-ave-Etot}. By comparing the peak energy in a rope during an event to the value before, we can determine the change in energy during an event. This is shown in \cref{tab:LP_E_breakdown}. In addition to the change in energy within a rope, we can look at the compressive work done on a rope. When a failed torus rope collapses, it moves into a region of larger external magnetic field, which causes the ropes to decrease in cross-sectional area. The change in volume of the wedge around the apex (See \citet{alt2022thesis} for a description of the wedge volume.) causes work to be done on the rope via \cref{eqn:E_conservation}. The total work done on a rope during this compression can be found by integrating the curve shown in \cref{fig:LP-compressive} over the time period of an event. 

The total work done on a rope can then be compared to the total change in energy to validate the ideal MHD model as a description of the energy inventory. When these two values are compared, it is seen that they agree to within 6\%. This is evidence that failed tori can be explained within the framework of ideal MHD without the need for non-ideal effects such as reconnection. Previous work has seen that around 7-8\% of the magnetic energy is dissipated during Taylor relaxation \citec{ji1995}. However, this value compares the energy dissipated to the total magnetic energy, while our 6\% agreement is relative to the measured energy change. This means that as long as the error in our thermal energy measurements is $\lesssim50\%$, our agreement still falls well below the expected energy dissipation.

In \cref{fig:LP-ave-Etot}, it can be seen that the total energy along with its components, magnetic and internal, drop by a factor of almost 2. This is a much larger change than would be expected in a device with fixed walls such as in \citet{ji1995}. This energy change is understood via the large change in the geometry of the system when a rope undergoes a failed torus event. During an event, the volume of a rope changes dramatically, causing a large change in energy. However, it is notable that even a large energy change of $\sim50\%$ can be explained through purely ideal MHD without resorting to non-ideal effects such a reconnection.

In this section, we have neglected the changes in the ion pressure, $P_i$. Due to the difficulty of measuring the ion temperature, we do not have any direct measurements of $P_i$ in our experiments. The characteristic timescale for energy transfer between electrons and ions is $\tau_{E,ei}\sim\unit[3]{\mu s}$ while the dynamic time of an event is $\tau_\ms{event}\sim\unit[2]{\mu s}$. Since these two values are comparable, the electrons should be able to transfer some energy to the ions during an event, but do not transfer energy fast enough to bring them into equilibrium at all times. Therefore, the actual situation is somewhere between $P_i=\const$ and $P_i=P_e$. The latter assumption would double the values of both the pressure compressive work and internal energy presented in \cref{tab:LP_E_breakdown}. This decreases the total work by $\unit[8\times10^{16}]{eV/cm}$ while increasing the total energy change by $\unit[12\times10^{16}]{eV/cm}$, bringing the two values closer to agreement. Therefore, the assumption of $P_i=\const$ does not weaken the main conclusion of this chapter that ideal MHD is sufficient to explain the failed torus events.

\section{Foot-point current measurements}
\label{sec:Rogowski-data}

In this section, we will investigate the changing current distribution at the foot points of the ropes measured via the Rogowski coils described in \cref{sec:Rogowski-description}. The total current varies slowly over the lifetime of a rope ($\tau_\ms{D}\sim\unit[150]{\mu s}$) which is also measured by the outermost Rogowski coil. Therefore, fast changes in the current measured by the other coils represent current redistribution rather than actual change in the current. From these measurements, we can observe how the current profile changes at the foot points of ropes during events. The change in current distribution at the foot points during failed torus events can be a further indication of hollowing current throughout the rope, strengthening the conclusions previously made based on measurements at the apex.

As a note on notation: we will define the current measured by one of the Rogowski coils with a symbol such as \RI{A1}. The subscript will start with either an ``A'' or a ``C'' to refer to coils on the anode or cathode side, respectively. The number can be any of $\{0,1,2\}$ and refers to specific coils printed into each PCB. For example, \RI{A0} refers to the current measured by the outermost coil on the anode side (which should be the same as the total plasma current), and \RI{C2} refers to the innermost coil on the cathode side. We will also use \RI{Xx} to refer to an arbitrary coil.

\subsection{Oscillating foot-point current envelope definition}
\label{sec:PCB-envelope-def}

\begin{figure}
    \centering
    \includegraphics[width=\linewidth]{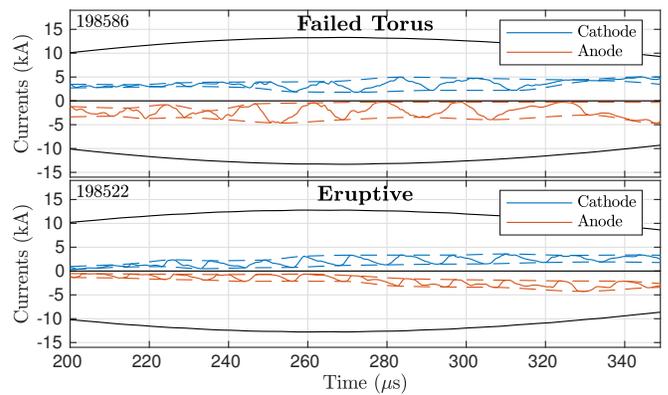}
    \caption{The currents measured by the middle Rogowski coils, \RI{A1} and \RI{C1}, are plotted in color compared to the total plasma current in black for both an example failed torus (top) and eruptive rope (bottom). For comparison with negative currents, $-I_\ms{p}$ is also shown. The envelopes around the oscillating current is shown with dashed lines. The time-averaged width of this envelope is used to determine how much the current distribution changes at each foot point.}
    \label{fig:PCB-envelope-def}
\end{figure}

In order to investigate the change in current at the foot points of the rope, we define an envelope around the oscillating currents. This is similar to the procedure used to define the instability amplitude of the apex height, $\langle\delta z\rangle/\xf$. An example of these envelopes based on \RI{X1} for two example ropes can be seen in \cref{fig:PCB-envelope-def}. Here the current in the middle Rogowski coils at both the anode and cathode are plotted for an example failed torus and eruptive rope. The total current passing through the electrodes is also plotted in black for reference. A normalized oscillation amplitude, \PCBenv{Xx}, is defined to measure the change in current normalized to the peak plasma current. Here $\langle\RI{Xx}\rangle$ is envelope width time-averaged over a period where the plasma current is within 80\% of its maximum value and $I_\ms{p,max}$ is the peak plasma current. 

\begin{figure}
    \centering
    \begin{subfigure}{\linewidth}
        \includegraphics[width=\linewidth]{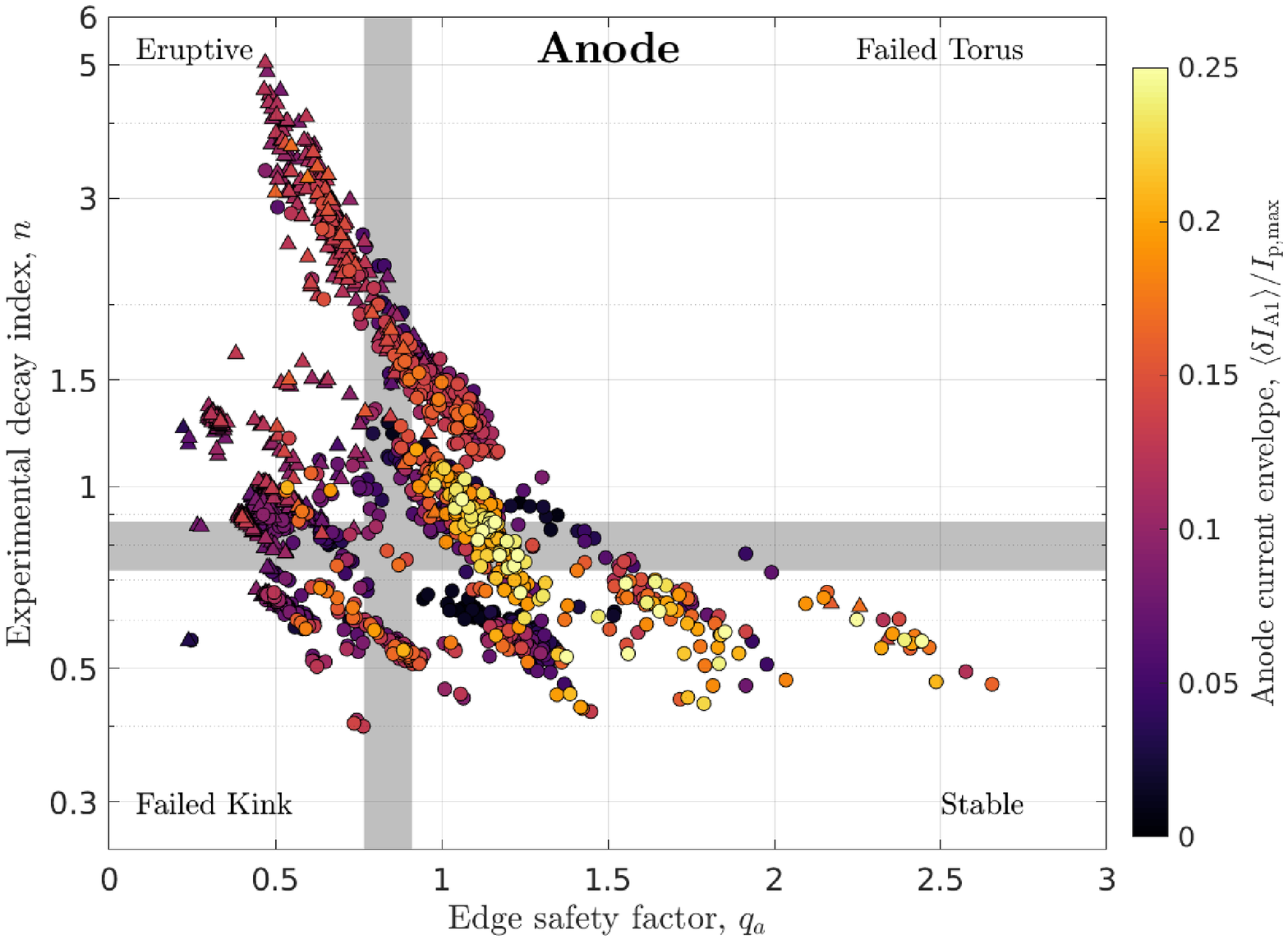}
        \phantomsubcaption\label{fig:Rogowski_all_anode}
    \end{subfigure}
    \begin{subfigure}{\linewidth}
        \includegraphics[width=\linewidth]{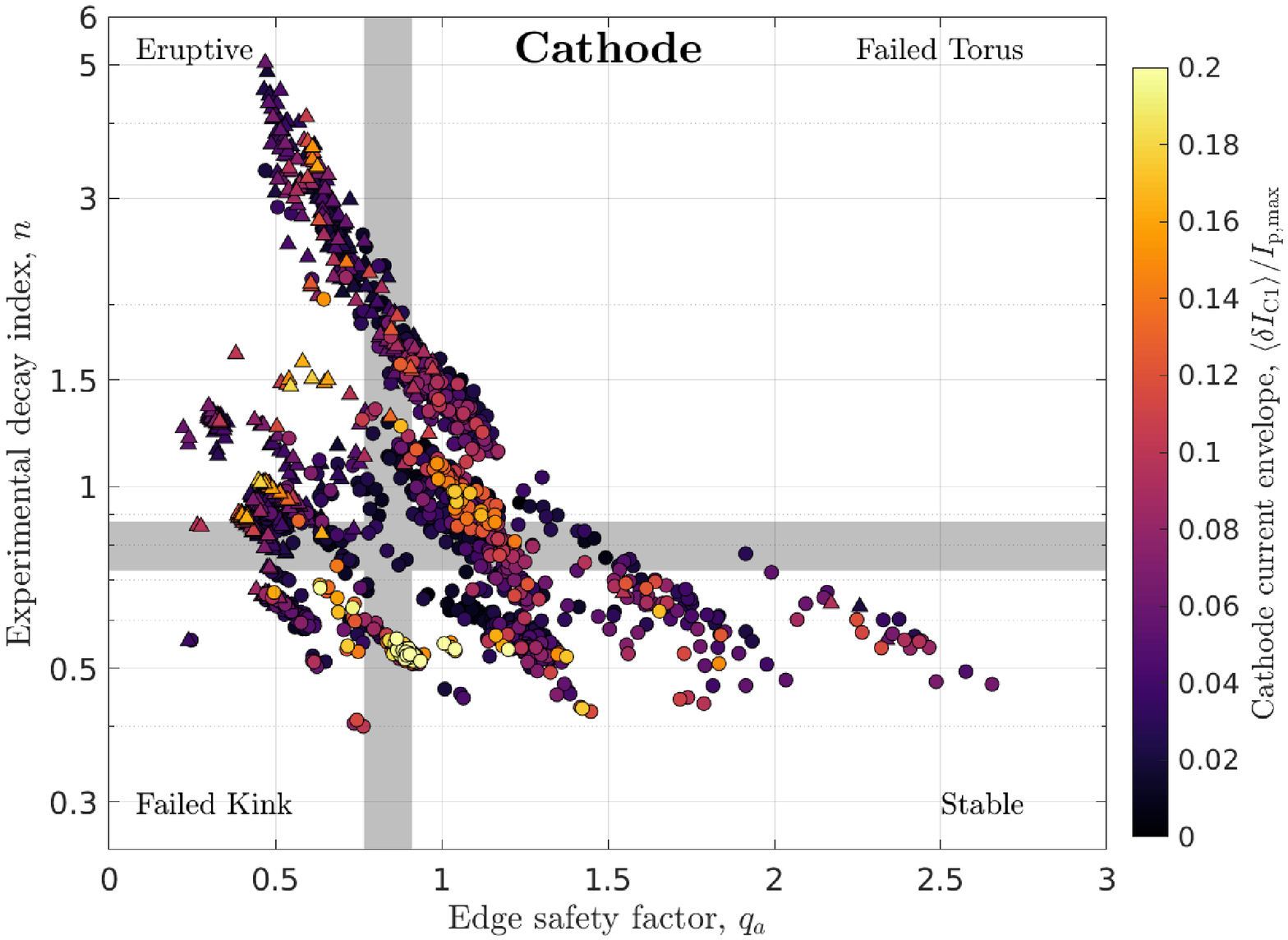}
        \phantomsubcaption\label{fig:Rogowski_all_cathode}
    \end{subfigure}
    \caption{The normalized foot-point current envelope of the middle Rogowski coils, \PCBenv{X1}, (in color) for both the anode (top) and the cathode (bottom).  As in \cref{fig:ns_vs_q}, each point is placed based on its value of the two ideal instability parameters, \qa{} and $n$ and shots that erupted are represented by triangles while non-eruptive shots are circles. The envelope sizes for the anode are significantly larger than those of the cathode due to the stronger line-tying there. The shots with the largest changes are in the failed torus regime rather than the most violently eruptive shots.}
    \label{fig:Rogowski_all}
\end{figure}

The values of \PCBenv{A1} and \PCBenv{C1} for each shot are shown in color in \cref{fig:Rogowski_all}. As in \cref{fig:ns_vs_q}, each shot is plotted based on its value of the stability parameters for the kink and torus instabilities, \qa{} and $n$ as described in \cref{sec:KI,sec:TI}, respectively. The results of almost 2,000 shots are shown. The electrons in the anode sheath are less mobile than those near the cathode, which causes more resistivity and thus weaker line-tying at the anode \citec{oz2011}. The stronger line-tying effects at the cathode cause the cathode envelopes to be smaller than the anode envelopes. To allow for easier visual inspection, the range represented by the color bar in \cref{fig:Rogowski_all_cathode} is slightly smaller than in \cref{fig:Rogowski_all_anode}. 

The largest values of \PCBenv{A1} are in the failed torus regime. Even though the eruptive ropes rise and erupt much more dramatically, the redistribution of current at the anode is not as significant. This is an indication that the current hollowing discussed in \cref{sec:FT-description} occurs throughout the rope in failed torus events, but is not significant in purely eruptive events. {The reason why there is not significant current hollowing in eruptive events is discussed in \cref{sec:kink-on-FT}.} Similarly, we see that \PCBenv{C1} is also larger in the failed torus regime than in the eruptive regime. However, the largest values of \PCBenv{C1} occur in kinked ropes. It is currently unclear why these ropes in particular had larger changes at the cathode than the anode. But since the focus of this work is not on the pure kink instability, we will focus on the top two quadrants. We now have evidence for current hollowing occurring during failed torus events at the apex and at both electrodes simultaneously.

\begin{figure*}
    \centering
    \includegraphics[width=\linewidth]{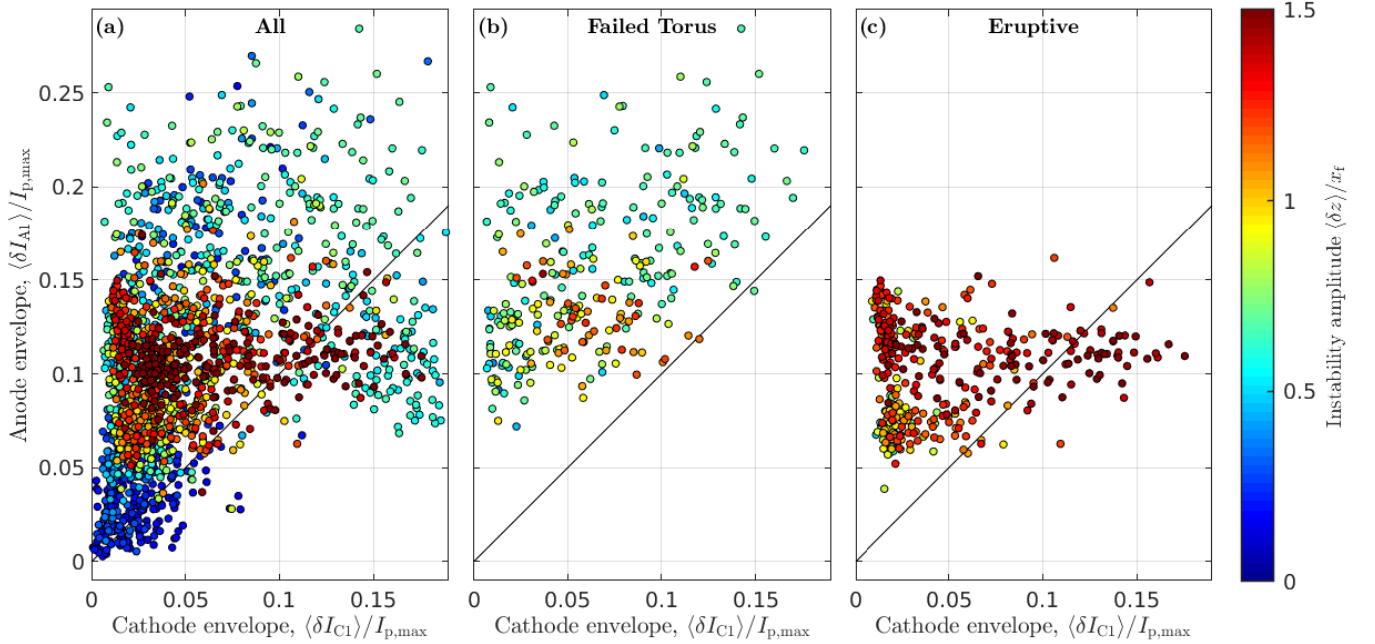}
    {\phantomsubcaption\label{fig:Rogowski_AvC_all}}
    {\phantomsubcaption\label{fig:Rogowski_AvC_FT_vs_erupt_FT}}
    {\phantomsubcaption\label{fig:Rogowski_AvC_FT_vs_erupt_erupt}}
    \caption{A comparison of the normalized foot-point current envelopes at the anode vs. the cathode for: \textbf{(a):} each shot, \textbf{(b):} failed torus ropes, and \textbf{(c):} eruptive ropes. The points in \textbf{(b)} and \textbf{(c)} are a subset of the points in \textbf{(a)}. The line $y=x$ has been added to aid comparison of the two quantities. The instability amplitude is also shown in color using the same scale as in \cref{fig:ns_vs_q}. Here we see that for the majority of ropes, $\PCBenv{A1}>\PCBenv{C1}$. The failed torus ropes have a generally larger \PCBenv{A1} while the spread on \PCBenv{C1} is similar.}
\end{figure*}

\subsection{Comparison of the anode and cathode foot-point current envelopes}

In \cref{fig:Rogowski_AvC_all} we have again plotted each of the shots shown in \cref{fig:Rogowski_all}, but now each shot is plotted based on its value of \PCBenv{C1} on the $x$-axis and \PCBenv{A1} on the $y$-axis, along with the instability amplitude, $\langle\delta z\rangle/\xf$, in color. This allows us to directly compare the changing current at the anode and cathode for each shot. From this plot, it is immediately obvious that $\PCBenv{A1}>\PCBenv{C1}$ for most shots, which is easily explained by the stronger line tying at the cathode. We also see that the most eruptive shots do not have the largest value of \PCBenv{A1}. Despite having the largest change in apex height, they do not have the most current redistribution at the foot points. The reason why eruptive ropes have less current redistribution than failed torus ropes is discussed in \cref{sec:kink-on-FT}.

In order to investigate this, it is helpful to separate the shots in \cref{fig:Rogowski_AvC_all} based on stability quadrant. In \cref{fig:Rogowski_AvC_FT_vs_erupt_FT,fig:Rogowski_AvC_FT_vs_erupt_erupt}, we recreate the same plot but with shots only in the failed torus regime (\cref{fig:Rogowski_AvC_FT_vs_erupt_FT}) or in the eruptive regime (\cref{fig:Rogowski_AvC_FT_vs_erupt_erupt}). Here, we see that the failed torus ropes have larger values of \PCBenv{A1} than the eruptive ropes. This indicates that there is more current redistribution occurring during failed torus events, which is indicative of the current hollowing that is occurring in these events. We also see that the ropes with $\PCBenv{A1}<\PCBenv{C1}$ are much more likely to be eruptive than failed torus. This could be due to the fact that new ropes are created during eruptive events and the profile change is new rope formation rather than current hollowing.

\subsection{Foot-point current redistribution during an event}

\begin{figure}
    \centering
    \includegraphics[width=\linewidth]{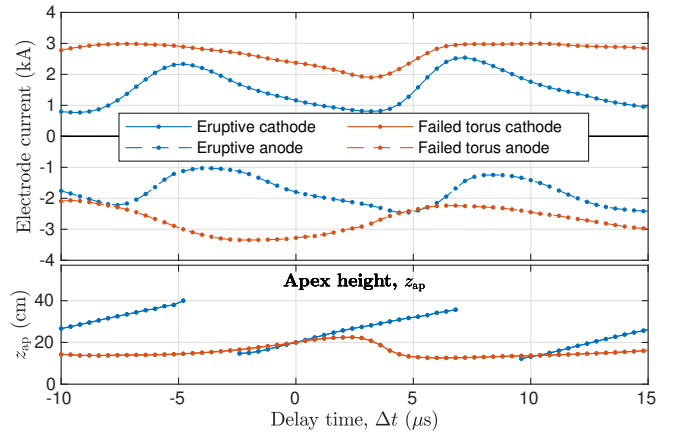}
    \caption{\textbf{Top:} The current measured by the middle Rogowski coils, \RI{X1}, on both the cathode (solid lines) and the anode (dashed lines) for both eruptive events (blue) and failed torus events (red). \textbf{Bottom:} The average apex height for each condition for comparison with the measured current. A larger current (in absolute value) represents a more peaked current distribution at the electrode.}
    \label{fig:PCB-during-event-middle-coil}
\end{figure}

We will now investigate the specific behavior of the measured foot-point current on the Rogowski coils during an average event. An average failed torus or eruptive event is created the same way as it was done in \cref{sec:LP-time}. The results of this for the middle coils are shown in \cref{fig:PCB-during-event-middle-coil}. The results of the smallest, center coils are qualitatively similar as the middle coils and can be seen in \citet{alt2022thesis}. This means that the redistribution of the current is not localized to a particular radius but occurs throughout the foot point. Since the middle Rogowski coil measures the current in both the center electrode segment and the middle ring, changes in \RI{X2} also cause changes in \RI{X1}. However, since the changes seen in \RI{X1} are about twice the size of the changes seen in \RI{X2}, we can still conclude that the current redistribution is occurring throughout the entire minor radius.

The change in currents during an average eruptive event in \cref{fig:PCB-during-event-middle-coil} look remarkably similar between the electrodes. However, since the nominal current to each electrode has a different sign, we should actually be focused on absolute value changes instead. In absolute value, \RI{A1} and \RI{C1} actually have opposite behavior. Leading up to an event, $|\RI{A1}|$ increases, (Implying that the current is becoming more peaked at the anode.) while $|\RI{C1}|$ decreases. (Implying that the current is becoming less peaked at the cathode.) The trend then reverses while a new rope is formed and the cycle begins again. This means that there is not a consistent current redistribution during these events and the field lines that are gaining current intersect the cathode and anode at different minor radii. 

The foot-point current change during an eruptive event can be contrasted with that of a failed torus event. During the rise, both $|\RI{A1}|$ and $|\RI{C1}|$
decrease. This means that there is less current going to the center two electrode segments\footnote{Since the total plasma current is constant, the decrease of the center currents also implies an increase to the current of the outer ring.}, implying that the current is becoming less peaked at both electrodes. Since the failed torus events are less coherent than the eruptive events (see \cref{sec:LP-time}), the return to the initial state is not as dramatic, but the current can still be seen returning to a more peaked state. This is the first evidence that the hollowing current previously seen in failed torus events occurs throughout the rope and is not localized to the apex where previous measurements were focused \citec{myers2015}. With this new evidence, we will develop a theory of how the current hollowing and subsequent toroidal flux enhancement can occur without violating the ideal MHD assumptions except in a small region around the electrodes.

\section{Toroidal flux change during failed torus events}
\label{sec:measured-T-flux-change}

\begin{figure}
    \centering
    \includegraphics[width=\linewidth]{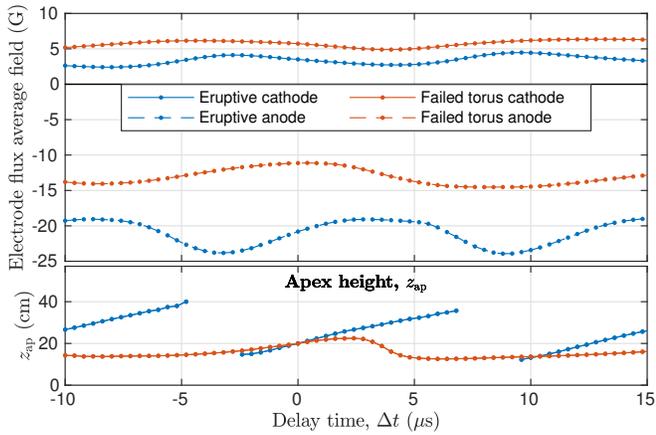}
    \caption{\textbf{Top:} The average toroidal magnetic field measured by the outer flux loop on both the cathode (solid lines) and the anode (dashed lines) for both eruptive events (blue) and failed torus events (red). \textbf{Bottom:} The average apex height for each condition for comparison with the measured flux. The majority of the toroidal flux in a rope is created on timescales that are too long to be measured by the flux loops. Therefore, the measurements here only show dynamic changes rather than absolute values.}
    \label{fig:PCB-flux-during-event}
\end{figure}

In addition to the measurements of the foot-point current distribution made possible by the PCBs, we can measure the toroidal flux through flux loops also printed into the boards. An example of such a measurement is shown in \cref{fig:PCB-flux-during-event} where the measured flux is normalized by the flux loop area to yield an average magnetic field, $\langle\delta B\rangle$. The majority of the toroidal flux in a rope is created by the external coils on timescales that are too long to be measured by the flux loops. Therefore, the measurements here only show dynamic changes rather than absolute values. In these plots we see that in both conditions, $\langle\delta B\rangle$ changes more on the anode than the cathode, consistent with the stronger line-tying there. The measured changes also lie within a relatively narrow range of $\langle\delta B\rangle\sim\unit[2-5]{G}$. Since the flux loops lie below the electrodes, which can be considered to be perfectly conducting on eruption time scales, the measured fluxes do not directly correspond to the toroidal flux in the ropes. In the rest of this section, we will devise a model to describe how these measurements would correspond to the changes in the ropes near the foot points.

\subsection{Toroidal flux change in a rising MFR}
\label{sec:flux-change-explanation}

\begin{figure}
    \centering
    \includegraphics[width=\linewidth]{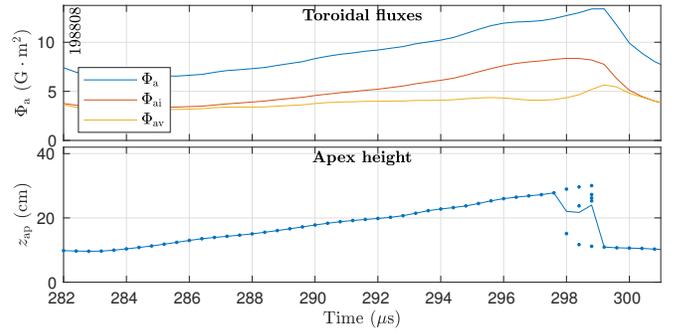}
    \caption{\textbf{Top:} The measured total, internal, and vacuum toroidal fluxes at the apex of a rope during an example failed torus event. The internal and total flux both increase during the rope's rise. \textbf{Bottom:} The apex height of the rope for comparison. During the failed torus event, the apex height is briefly multi-valued.}
    \label{fig:apex-flux-during-event}
\end{figure}

\newcommand{\figW}{0.5}
\begin{figure*}
    \centering
    \begin{subfigure}{\figW\linewidth}
        \includegraphics[width=\linewidth]{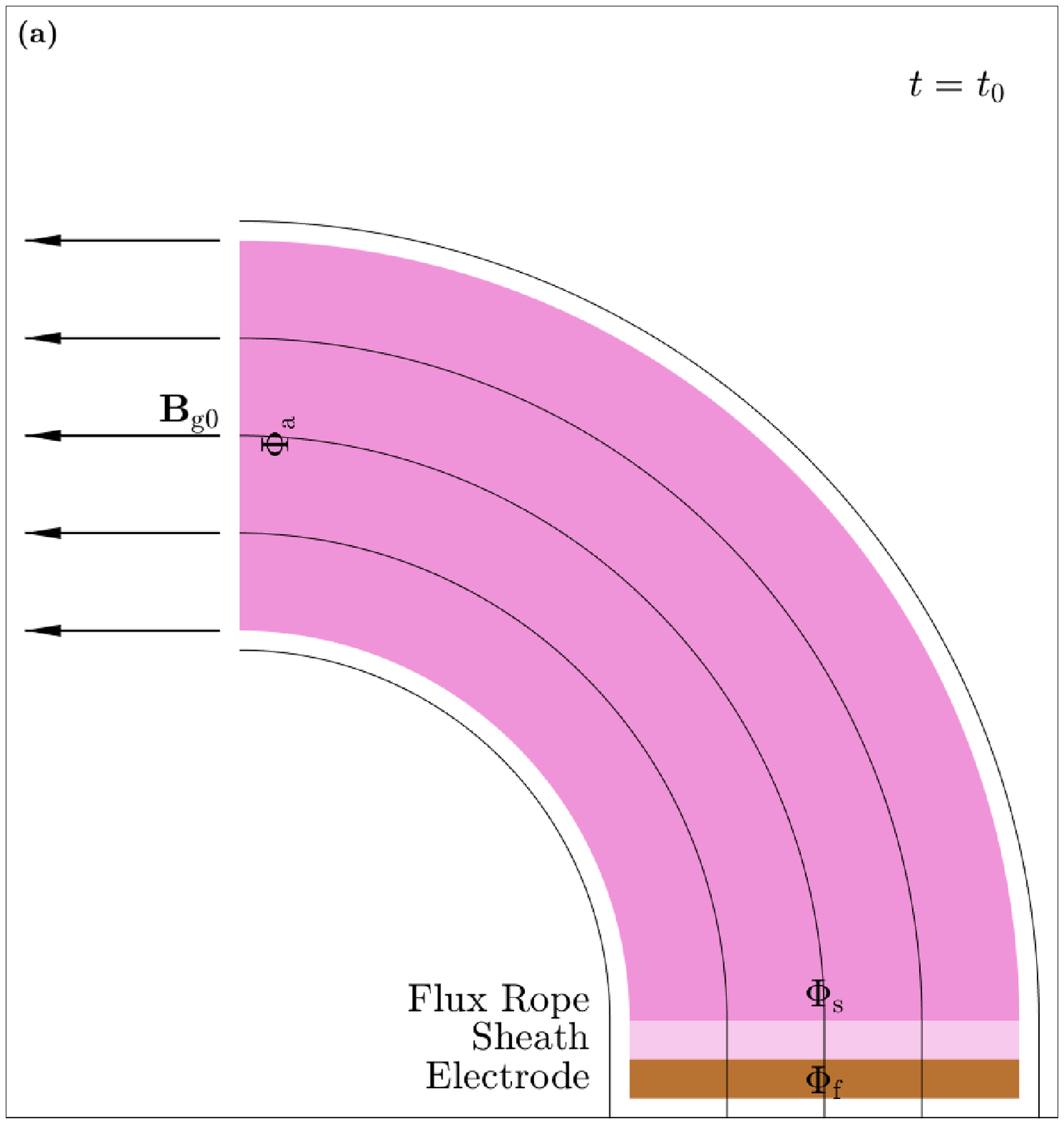}
        \phantomsubcaption\label{fig:flux-inc-explanation-a}
    \end{subfigure}%
    \begin{subfigure}{\figW\linewidth}
        \includegraphics[width=\linewidth]{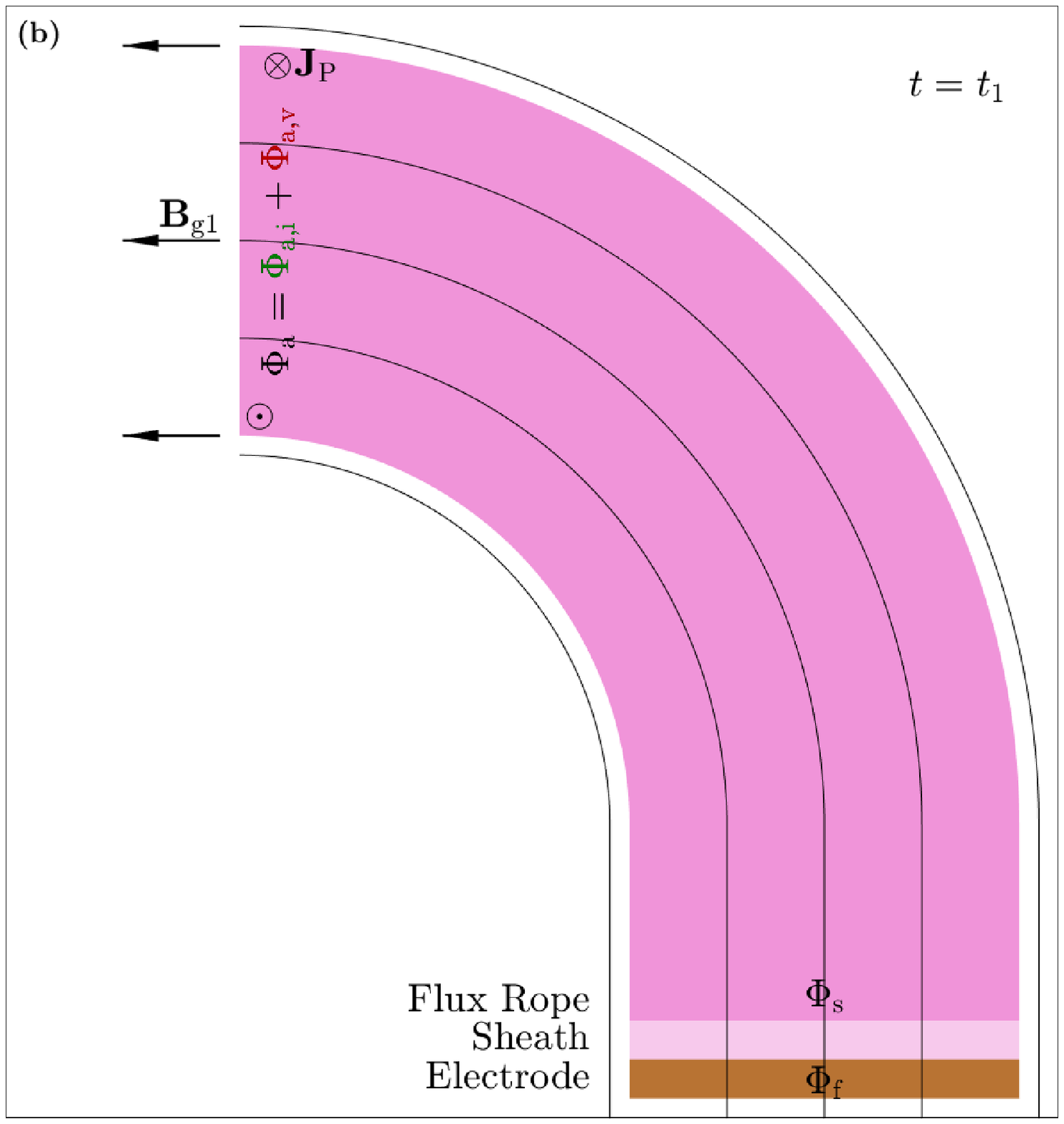}
        \phantomsubcaption\label{fig:flux-inc-explanation-b}
    \end{subfigure}
    \begin{subfigure}{\figW\linewidth}
        \includegraphics[width=\linewidth]{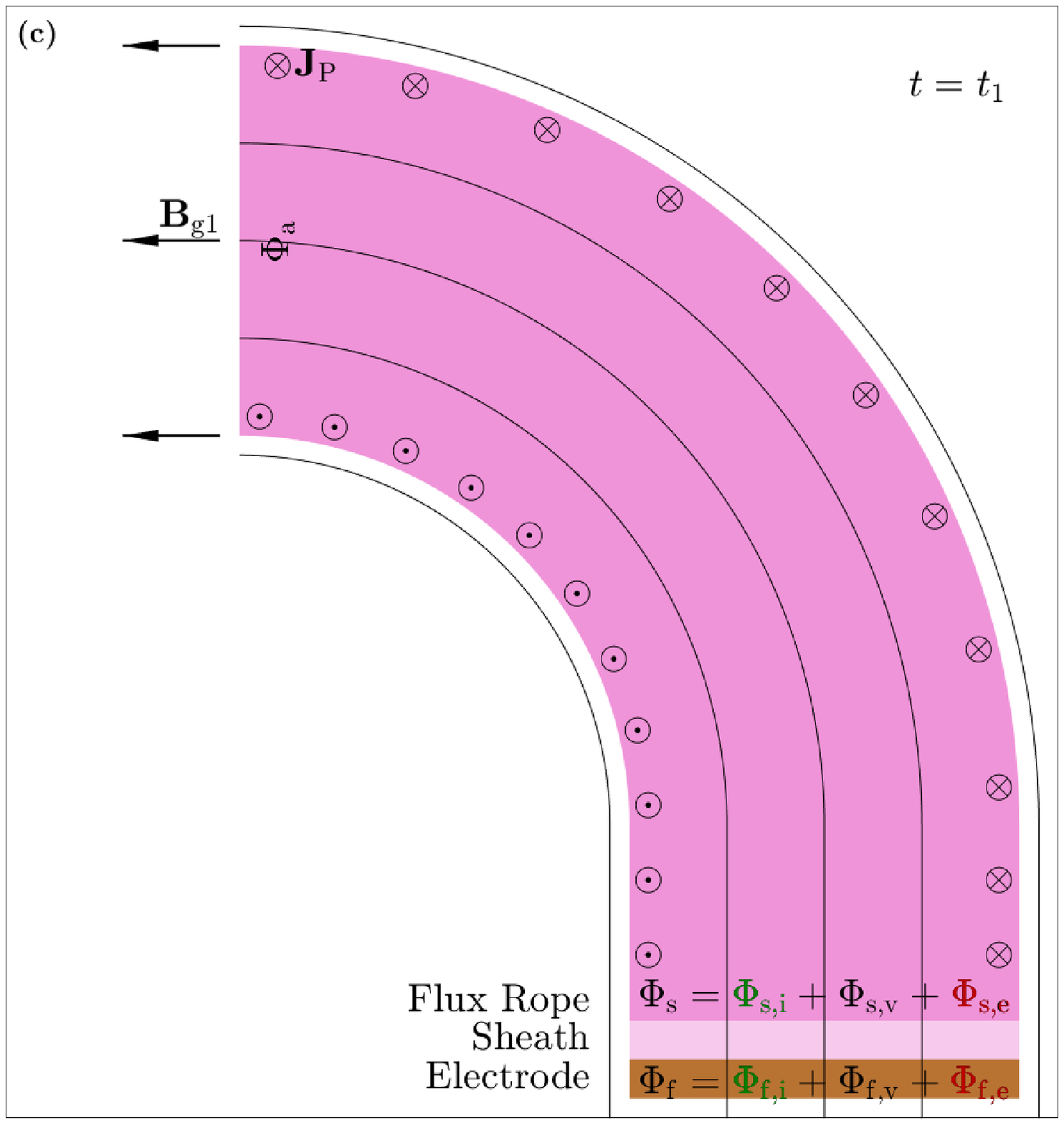}
        \phantomsubcaption\label{fig:flux-inc-explanation-c}
    \end{subfigure}%
    \begin{subfigure}{\figW\linewidth}
        \includegraphics[width=\linewidth]{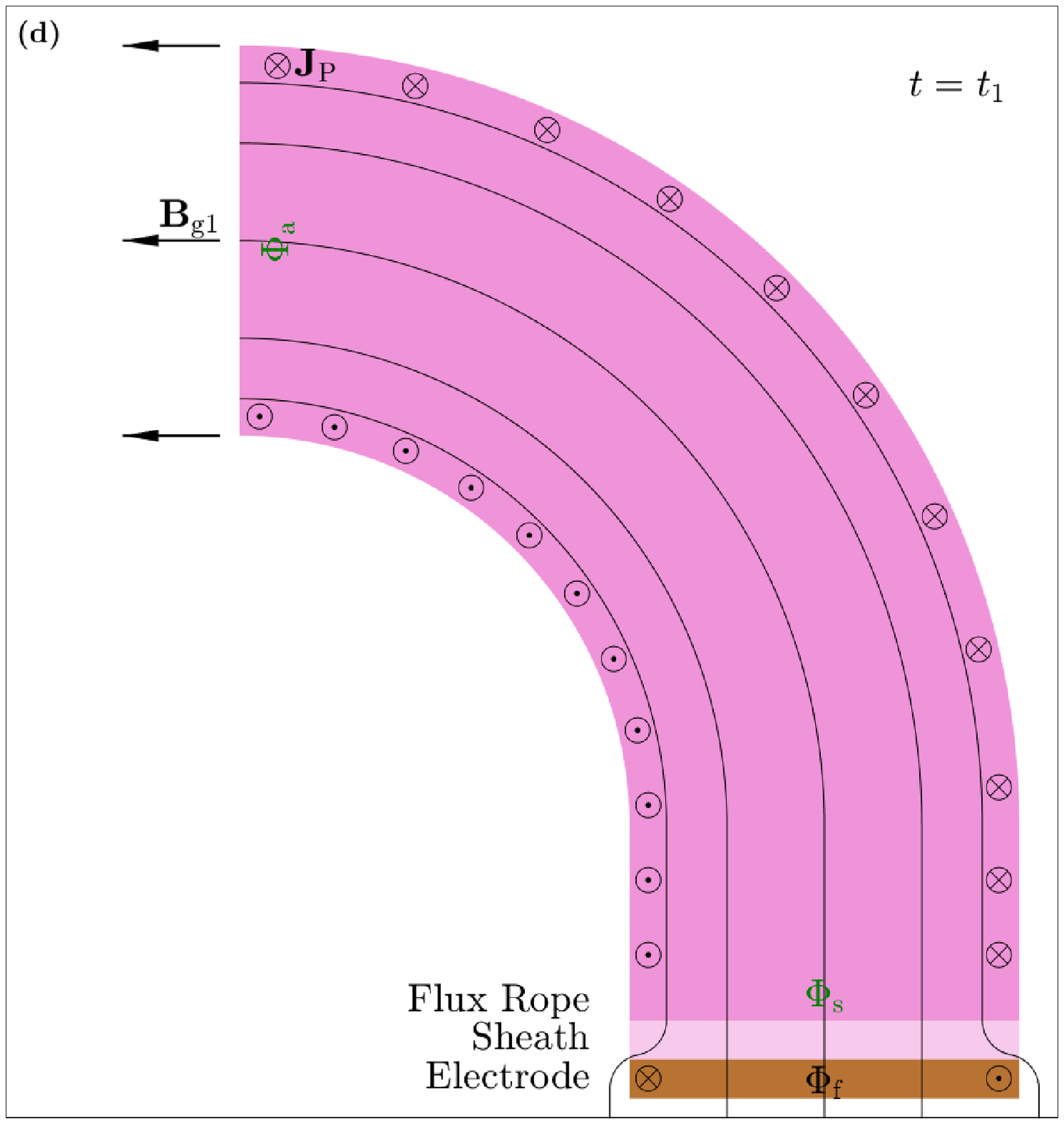}
        \phantomsubcaption\label{fig:flux-inc-explanation-d}
    \end{subfigure}%
    \caption{Cartoon of the changing toroidal fluxes in a rising rope. The fluxes are considered at three locations, the apex, the top of a non-ideal sheath layer, and at the electrode surface. Terms that increase are marked in green while terms that decrease are marked in red. \textbf{(a):} The initial state of the rope with guide field, \Bg[0]. \textbf{(b):} The rope rises into a region with smaller guide field, \Bg[1], and so \JP{} is enhanced to increase \fluxT{a,v} so that toroidal flux is conserved. \textbf{(c):} Since the current follows field lines, \JP{} is enhanced throughout the rope, increasing \fluxT{s,i}. \textbf{(d):} Eddy currents are generated in the electrode to conserve \fluxT{f}, this bends field lines into the rope and increases \fluxT{s} and therefore \fluxT{a}.}
    \label{fig:flux-inc-explanation}
\end{figure*}

The toroidal flux in a rope has been seen to increase during failed torus events as shown in \cref{fig:apex-flux-during-event}. The cartoon in \cref{fig:flux-inc-explanation} illustrates a series of events wherein this can happen. In \cref{fig:flux-inc-explanation-a}, we have the initial state of our rope. The background guide field is $B_\ms{g0}$ and we will discuss the toroidal flux in three locations: at the apex, \fluxT{a}, at the sheath edge \fluxT{s}, and at the electrode surface, \fluxT{f}. The surface for \fluxT{s} is the edge of a non-ideal layer at the base of the rope where non-ideal MHD effects cannot be ignored, such as a Debye sheath. Other potential non-ideal layers are discussed in \cref{sec:non-ideal-layers}. Since the rope is line-tied at the electrode, \fluxT{f} is a constant. The three fluxes can also be broken up based on contributing fields. We will separately consider the internal flux created by currents in the rope, the vacuum flux created by the external magnetic field coils, and the eddy flux created by eddy currents in the copper electrode that exist to maintain the line-tied condition. These sources will be denoted with the subscripts ``i'', ``v'', and ``e'', respectively. The fluxes can then be written as
\begin{subequations}
\begin{align}
    \fluxT{a} &= \fluxT{a,i} + \fluxT{a,v} ,\\
    \fluxT{s} &= \fluxT{s,i} + \fluxT{s,v} + \fluxT{s,e} ,\\
    \fluxT{f} &= \fluxT{f,i} + \fluxT{f,v} + \fluxT{f,e} ,
\end{align}
\end{subequations}
where the eddy contribution to \fluxT{a} has been ignored since the eddy fields should be localized around the electrode. We can consider the initial values of \fluxT{s,e} and \fluxT{f,e} to be zero by including any initial eddy current fields in the vacuum fluxes.

In \cref{fig:flux-inc-explanation-b}, the rope has risen to a height with reduced guide field, $B_\ms{g1}$. This causes \fluxT{a,v} to decrease (indicated by the red text) and therefore \fluxT{a,i} must increase (indicated by the green text) to conserve \fluxT{a}. This is accomplished by increasing the poloidal current, \JP, at the apex. Since the plasma is force-free, the current must flow along field lines. Therefore, an increase in \JP{} at the apex leads to an increase along the entire rope, leading to \cref{fig:flux-inc-explanation-c}. The increase in \JP{} leads to an increase in both \fluxT{s,i} and \fluxT{f,i}. Since \fluxT{f} must be exactly conserved, this means that \fluxT{f,e} and \fluxT{s,e} must become negative. The fields due to the eddy currents are highly localized to the electrode, therefore $|\fluxT{s,e}|<|\fluxT{f,e}|$. This means that the total flux, \fluxT{s}, will increase as in \cref{fig:flux-inc-explanation-d}. The flux increase is caused by field line bending and finite $\bb{J}\btimes\bb{B}$ is allowed in the non-ideal layer.\footnote{The amount of field line bending allowed is examined via a simplified model in \cref{sec:flux-around-electrode}.} The increase in \fluxT{s} also leads to an increase in \fluxT{a} since ideal MHD can be applied to the rest of the rope. 

The increase in toroidal flux throughout the rope does not violate the frozen-in condition of ideal MHD because the new flux comes from incorporating new flux surfaces into the rope rather than changing the flux in a fluid element. The new field lines do not terminate on either electrode. However, they are able to carry plasma current because this current is able to cross field lines in the non-ideal layer and therefore still reach the electrode. 

The increase in \fluxT{a,i} requires an increase in \JP{} away from the center of the rope. This causes the current profile to be less peaked in the center, the extreme case of which is a hollowing of the current where it is largest at the edges of the rope as seen in \cref{fig:FT_time_slice}. While this current profile is only required of \JP{}, the profile must also be mirrored in \JT{} because the rope should be force free, i.e. the current must flow along field lines.

\subsection{Fields in a straight MFR with decreasing guide field}
\label{sec:straight-rope-flux}

One assumption made in \cref{sec:flux-change-explanation} is that \fluxT{a,v} decreases when the rope enters a region with lower guide field. However, it is not immediately obvious that this must be the case. One could imagine a rope entering a low \Bg{} region and expanding \textit{proportionally} to maintain \fluxT{a,v} rather than enhance the toroidal field. In this section, we will argue that this cannot be the case if flux and energy are to be conserved. We will do this using a simplified 1D straight flux rope model.

Consider a straight flux rope with radius, $a_0$, constant guide field, \Bg[0], and constant internal toroidal field, \Bti[0], so that its total toroidal field is $\BT[0]=\Bti[0]+\Bg[0]$. Then the toroidal flux and magnetic energy are given by
\begin{subequations}
\begin{align}
    \fluxT{T0} &= \pi a_0^2\left(\Bti[0] +\Bg[0]\right) ,\\
    W_0 &= \frac{\pi a_0^2}{2\mu_0}\left(\Bti[0] +\Bg[0]\right)^2 .
\end{align}
\end{subequations}
Now suppose we reduce the guide field to \Bg[1], where $\Bg[1]<\Bg[0]$. We will then determine the values for the other resulting rope parameters, $a_1$ and \Bti[1]. Using toroidal flux conservation, $\fluxT{0}=\fluxT{1}$, we have,
\begin{equation}
    \pi a_0^2(\Bti[0] +\Bg[0]) = \pi a_1^2(\Bti[1] +\Bg[1]) .
\end{equation}
Due to energy conservation, the change in magnetic energy must be equal to the work done by the rope on the external (guide) fields.\footnote{We have assumed zero $\beta$ so the internal energy cannot change and have neglected changes in the poloidal magnetic field energy. A poloidal energy term can be added to the LHS of \cref{eqn:straight-MFR-work}, however this term is generally negative for an arched MFR and increases the resulting \Bti[1].} The compressive work done by the external fields is caused by the magnetic pressure,
\begin{equation}
    W_\ms{comp,g}=\int P_\ms{comp,g} \rmd t = \int \Bg^2\D{t}{V} \rmd t,
\end{equation}
where $V\propto a^2$ is the volume of the rope. %
This yields\footnote{In general, $W_\ms{comp,g}$ will depend on the form of $\Bg(V)$. If one assumes an ansatz of $\Bg=\alpha V^{-n}$ for some $\alpha$ and $n>1$, then the RHS of \cref{eqn:straight-MFR-work-b} is multiplied by $1/(2n-1)$. This changes the specific values but not the general conclusions of this section. The condition of $n>1$ is equivalent to the rope expanding slower than a vacuum rope with $\Bti=0$.}
\begin{subequations}\label{eqn:straight-MFR-work}
\begin{align}
    W_1-W_0 &= W_\ms{comp,g} ,\\ 
    a_1^2(\Bti[1] +\Bg[1])^2 -a_0^2(\Bti[0] +\Bg[0])^2  &= a_1^2\Bg[1]^2 -a_0^2\Bg[0]^2 . \label{eqn:straight-MFR-work-b}
\end{align}
\end{subequations}
These two equations can be solved for $a_1$ and \Bti[1],
\newcommand{\discriminant}{\ensuremath{\Bti[0]^2(\Bti[0]+2)^2 +4\Bg[1]^2(\Bti[0]+1)^2 }}
\begin{subequations}
\begin{align}
    \left(\frac{a_1}{a_0}\right)^2 &= \frac{-\Bti[0]^2-2\Bti[0] \pm\sqrt{D} }{2\Bg[1]^2} ,\\
    \Bti[1]&= \frac{\Bti[0]^2-2\Bg[1]\Bti[0] -2\Bg[1]+2\Bti[0] \pm\sqrt{D} }{2(\Bti[0]+1)} ,
\end{align}
\end{subequations}
where the magnetic fields have been normalized to \Bg[0] and $D$ is defined as 
\begin{equation}
    D=\discriminant .
\end{equation}
We are only interested in the positive root of these equations in order to maintain $a_1^2>0$. It can be seen that there are only two free parameters in these equations, \Bg[1] and \Bti[0].

\begin{figure}
    \centering
    \begin{subfigure}{.8\linewidth}
        \includegraphics[width=\linewidth]{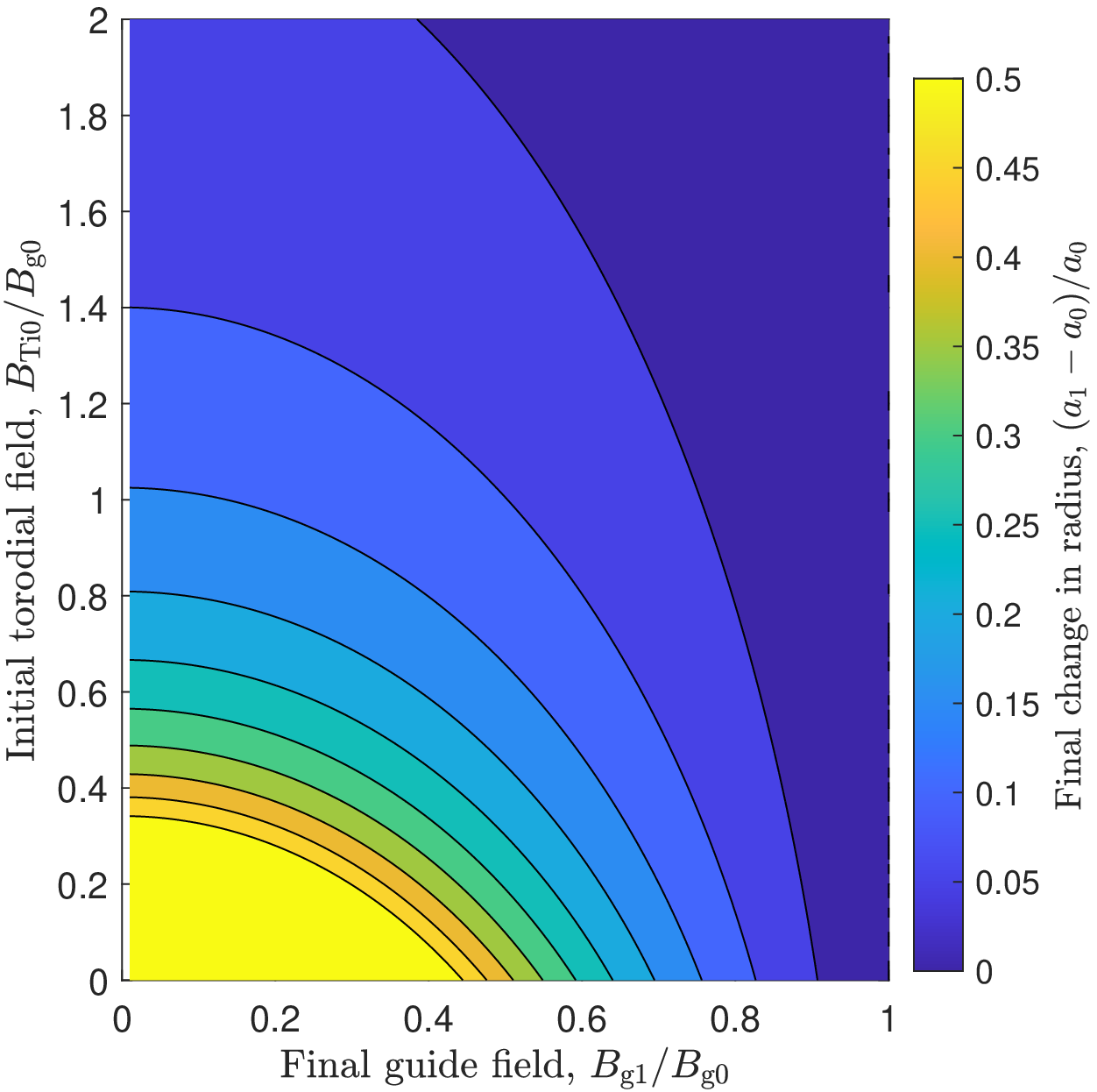}
        \phantomsubcaption\label{fig:straight-MFR-contours-a1}
    \end{subfigure}
    \begin{subfigure}{.8\linewidth}
        \includegraphics[width=\linewidth]{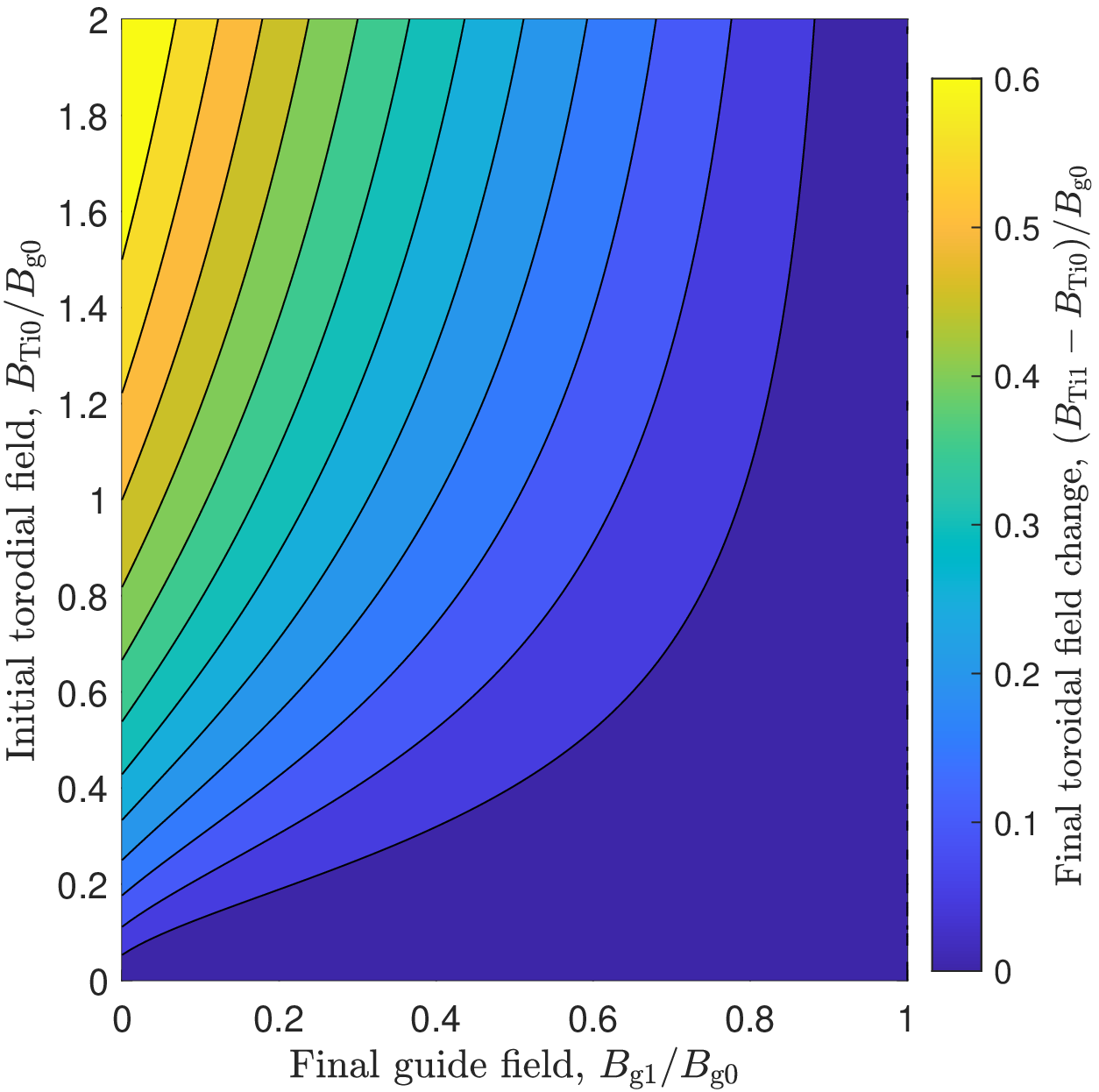}
        \phantomsubcaption\label{fig:straight-MFR-contours-Bti1}
    \end{subfigure} 
    \caption{Contour plots of the normalized change in the minor radius (left) and internal toroidal field (right) in a straight flux rope with decreasing guide field. When quantities are normalized to $a_0$ and \Bg[0], there are only two free parameters left, \Bg[1] and \Bti[0] which are represented on the $x$ and $y$-axes, respectively. Since both quantities are non-negative, both the minor radius and internal toroidal field increase when the guide field is decreased.}
    \label{fig:straight-MFR-contours}
\end{figure}

Contour plots of $a_1/a_0-1$ and $(\Bti[1]-\Bti[0])/\Bg[0]$ are presented in \cref{fig:straight-MFR-contours}. It can be seen in these plots that both of these quantities are nonnegative for all values of \Bg[1] and \Bti[0]. Therefore, the rope expands and the internal toroidal field is enhanced by the decrease in external guide field. While this simplified model without toroidal effects cannot be effectively used to quantitatively describe the changing vacuum flux during the rise of an arched rope, it does indicate that we should expect \fluxT{a,v} to decrease as a rope rises into areas of smaller \Bg{}. Since this simplified model does not include any effects of the field profile, (i.e. it assumes constant \BT) it cannot accurately predict the enhancement of the tension force. A comparison to experimental profiles during an event with constant flux similar to this model are presented in \cref{sec:flux-indexed-values}.

\subsection{Toroidal flux enhancement by field-line bending around the electrodes}
\label{sec:flux-around-electrode}

\begin{figure}
    \centering
    \begin{subfigure}{.49\linewidth}
        \includegraphics[width=\linewidth]{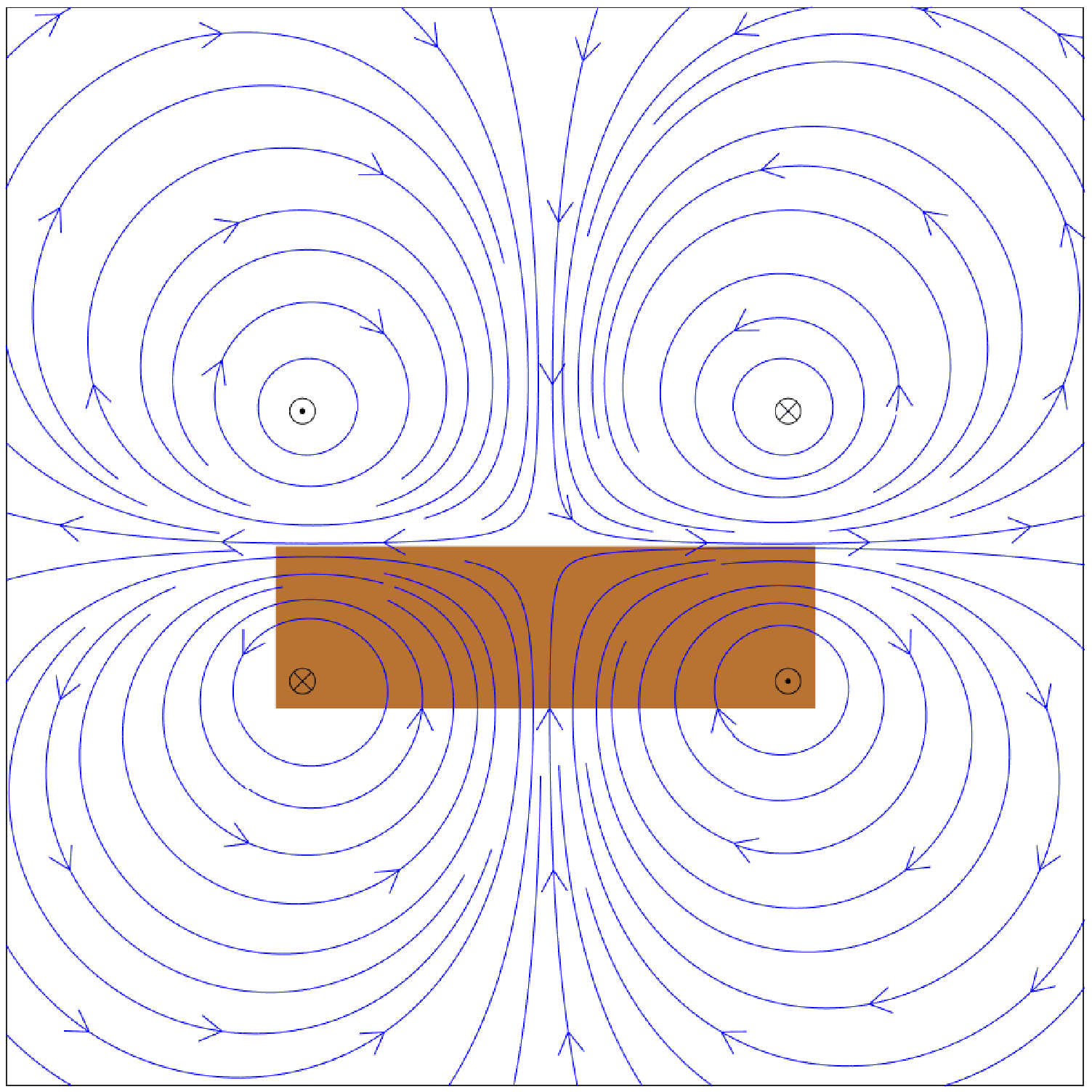}
        \phantomsubcaption\label{fig:current-loop-isolated}
    \end{subfigure}%
    \hfill
    \begin{subfigure}{.49\linewidth}
        \includegraphics[width=\linewidth]{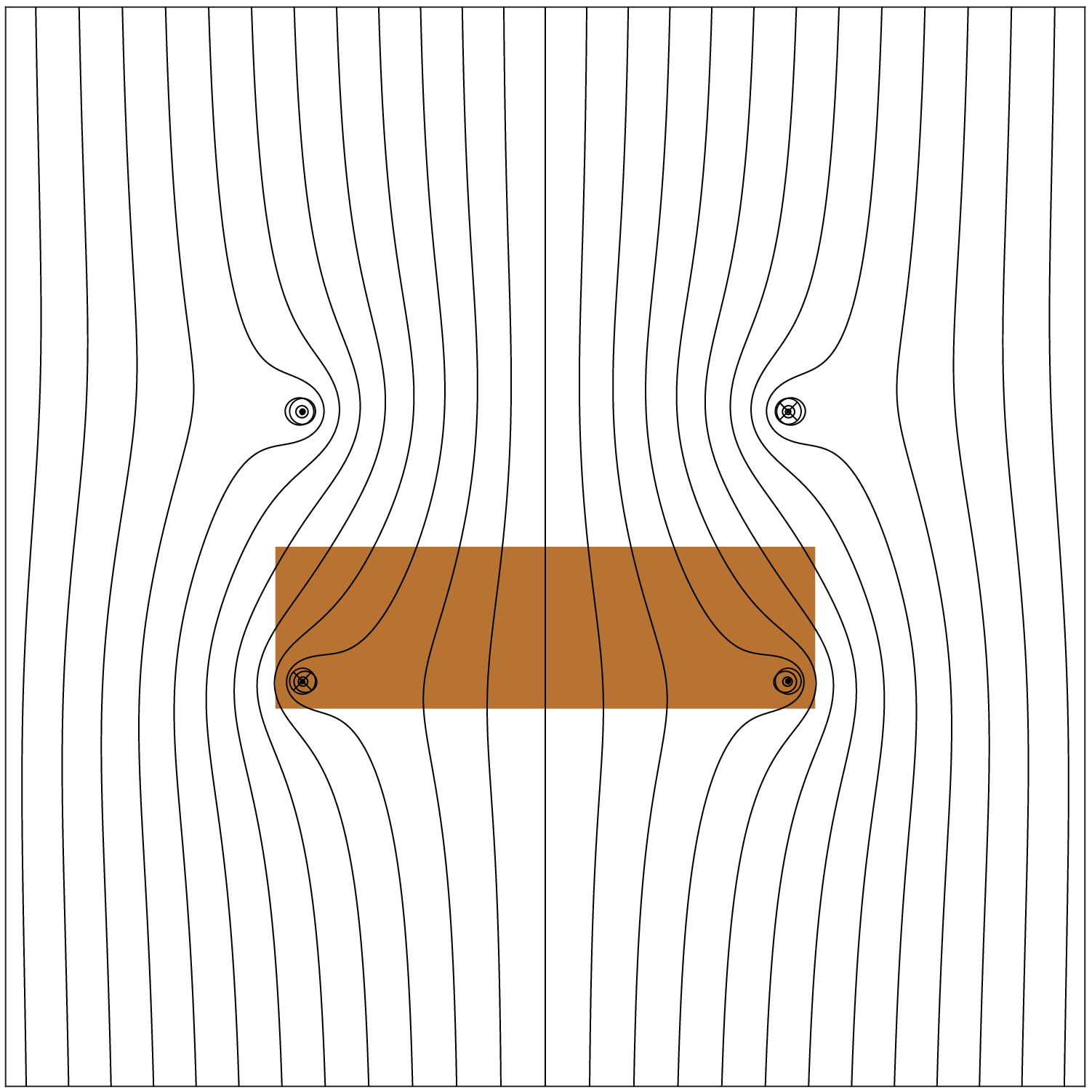}
        \phantomsubcaption\label{fig:current-loop-full-lines}
    \end{subfigure}
    \caption{The field lines of a current loop located $\Delta z$ above the electrodes (shown in brown) along with an image current to maintain the frozen-in condition at $z=0$. \textbf{Left:} The field lines of just the current loops. \textbf{Right:} The current loops embedded in a background field in the $z$-direction. The electrode thickness and the sheath thickness have both been exaggerated for clarity.}
    \label{fig:current-loop-field}
\end{figure}

In order to better study the field-line bending occurring in \cref{fig:flux-inc-explanation}, a simple model of enhanced poloidal current is devised. We will model the current as a infinitesimal current loop with current $I_\ms{loop}$ of radius $a$ located at a height $z=\Delta z$ above the electrode. This height can be taken as the edge of the non-ideal layer where the plasma current is able to cross field lines. In order to maintain the frozen-in condition at the electrode surface, $\Delta B(z=0)=0$, a mirror current loop of current $-I_\ms{loop}$ is located at $z=-\Delta z$. The field lines of this configuration are shown in \cref{fig:current-loop-isolated}. \cref{fig:current-loop-field} is not to scale. In particular, the ratio $\Delta z/a$ has been increased to ease visibility of the field lines in the non-ideal layer. 

These field lines are only meant to represent the field due to the enhanced poloidal current, and so \cref{fig:current-loop-full-lines} shows the field lines of these current loops embedded in a straight field representing the guide field. Here one can see that field lines that would not be in the rope without the loop currents have been bent into the rope. In reality the the enhanced \JP{} should exist along the entire length of the rope rather than being localized to the base of the rope. While this simplified model can reasonably describe the field-line bending in the non-ideal layer, it cannot be used above the current loop at $z=\Delta z$. So, while the additional lines are later bent back out of the rope, in reality we expect that the enhanced toroidal flux will be frozen into the rope

\begin{figure}
    \centering
    \includegraphics[width=\linewidth]{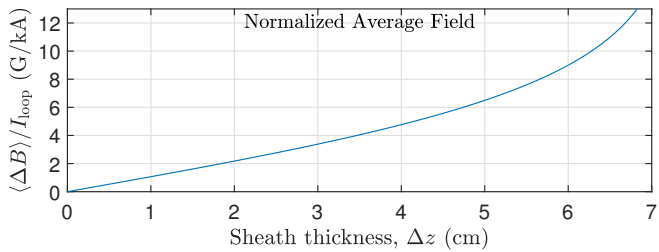}
    \caption{The average field at the flux loops calculated from the current loop model and using the actual experimental geometry. The average field, $\langle\Delta B\rangle$, is normalized to the loop current, $I_\ms{loop}$. The sheath thickness, $\Delta z$, sets the location of the current loop above the electrode. For $\Delta z\ll a_0$ this relation is approximately linear, where $a_0=\unit[7]{cm}$ is the electrode radius.}
    \label{fig:current-loop-flux}
\end{figure}

Using the experimental values for $a$ and the flux loop location in this model, we can predict the expected measured flux change (Shown as an average field change, $\langle\Delta B\rangle$.) as a function of $I_\ms{loop}$ and $\Delta z$, as shown in \cref{fig:current-loop-flux}. For small separations where $\Delta z\ll a$, $\langle\Delta B\rangle$ is approximately linear and is described by
\begin{equation}
    \langle\Delta B\rangle ~ (\unit{G}) \approx -1.06 ~ I_\ms{loop} (\unit{kA}) ~ \Delta z ~ (\unit{cm}) ,
\end{equation}
where $\langle\Delta B\rangle$ is measured in \unit{G}, $I_\ms{loop}$ is measured in \unit{kA}, and $\Delta z$ is measured in \unit{cm}. As a rough estimate, we expect that $I_\ms{loop}\lesssim \unit[10]{kA}$ since it should not be significantly larger than the toroidal current, $I_\ms{P}\approx\unit[12]{kA}$. As seen in \cref{sec:measured-T-flux-change}, the measured change in flux by the flux loops is $\langle\Delta B\rangle\sim\unit[2-5]{G}$. Based on these numbers, we expect that the non-ideal layer has a height of $\Delta z\sim \unit[0.2]{cm}$. %
While this simplified model can only be trusted for an order-of-magnitude estimate of $\Delta z$, we can still compare the expected non-ideal based on the plasma parameters to this value.

\subsection{Possible non-ideal layers between an MFR and electrode}
\label{sec:non-ideal-layers}

We expect the non-ideal layer between the electrodes and the base of the flux rope to have a width of around $\Delta z\sim\unit[0.2]{cm}$. Three potential layers will be considered which are summarized in \cref{tab:non-ideal-layers}. The simplest potential layer is a Debye sheath with a width based on the Debye length, $\lambda_\ms{D}=\sqrt{{\epsilon_0 T}/{(n e^2)}}$, where $\epsilon_0$ is the permittivity of free space. However, based on the experimental parameters, $\lambda_\ms{D}\approx\unit[2\times10^{-4}]{cm}$ which is far too small to explain the expected non-ideal layer. 

\renewcommand{\arraystretch}{1.0}
\begin{table}
    \centering
    \begin{tabular}{lcc}
         \toprule
         Layer type   & Experimental value & Solar value \\
         \midrule
         Debye sheath   &  $\unit[2\times 10^{-4}]{cm}$ & $\unit[1\times 10^{-4}]{cm}$\\
         $\lambda_\ms{D}=\sqrt{\epsilon_0 T/(n e^2)}$ & &\\ \midrule
         Hartmann layer    & \unit[0.35]{cm} & $\unit[0.086]{cm}$\\
         $\delta_\ms{Ha}={\sqrt{m_i n\nu\eta}}/{B}$ && \\ \midrule
         Resistive layer    & \unit[0.052]{cm}  & $\unit[2.1]{cm}$\\
         $\delta_\ms{S}={\eta\sqrt{\mu_0 m_i n}}/{B}$ && \\ 
         \bottomrule
    \end{tabular}
    \caption{A comparison of the potential non-ideal layers considered for the region between the electrode and the ideal rope plasma. Values for both the experimental conditions and the photosphere are presented. Sources of anomalous resistivity and viscosity have not been considered.}
    \label{tab:non-ideal-layers}
\end{table}

The Debye length describes the scale under which the quasi-neutrality approximation holds. However, we can consider the finite viscosity, $\nu$, and resistivity, $\eta$, in our plasma to find a characteristic scale length under which the ideal MHD approximations are violated. One example of this is the Hartmann layer where magnetic forces are balanced by resistive and viscus forces as described by the Hartmann number $\ms{Ha}={B \delta}/{\sqrt{m_i n\nu\eta}}$, 
where $\delta$ is a characteristic scale length \citec{hartmann1937,lingwood1999}. In Hartmann flow of a magnetized plasma over a stationary surface, a characteristic length scale over which $\bb{J}\btimes\bb{B}$ forces are balanced by viscus and resistive forces is given by $\delta_\ms{Ha}={\sqrt{m_i n\nu\eta}}/{B}$. %
Based on our experimental plasma parameters, this length scale is $\delta_\ms{Ha}\approx\unit[0.35]{cm}$, which is similar to the expected $\Delta z$.

In addition to $\delta_\ms{Ha}$, we could consider a length scale where magnetic forces are balanced entirely by resistive forces. This scale can be found by considering the Lundquist number, $\ms{S}=\mu_0 v_\ms{A} \delta/\eta$. %
Similar to how $\delta_\ms{Ha}$ is the length scale under which $\ms{Ha}=1$, we can find $\delta_\ms{S}$ as a length scale where $\ms{S}=1$, $\delta_\ms{S}={\eta\sqrt{\mu_0 m_i n}}/{B}$. %
Based on our experimental plasma parameters, this length scale is $\delta_\ms{S}\approx\unit[0.052]{cm}$. This is slightly too small to explain the expected $\Delta z$, so we expect the Hartmann layer to be the significant cause of the non-ideal layer.

The solar length scales for these non-ideal layers is also presented in \cref{tab:non-ideal-layers}. Compared to the typical length scales of solar MFRs, we see that these layers are all too small to be able to explain any significant flux change in a rope. However, we have only considered classical Spitzer resistivity and viscosity in these calculations. The addition of anomalous sources of resistivity and viscosity may increase the solar values significantly.

\section{Experimentally measured changes in the toroidal field profile}

\label{sec:flux-indexed-values}

The simple 1D straight flux rope model presented in \cref{sec:straight-rope-flux} can help explain the increase in \fluxT{a,i}. However, it fails to accurately predict the value of the enhanced tension force, \Ft, because it assumes the fields are constant in the rope. The \Ft{} change we are interested in depends on the fields via \cref{eqn:FTpara}. Since the relevant quantity is the average squared toroidal field, $\langle \BT^2\rangle$, the exact profile is necessary to predict the changes in \Ft.

\begin{figure}
    \centering
    \includegraphics[width=\linewidth]{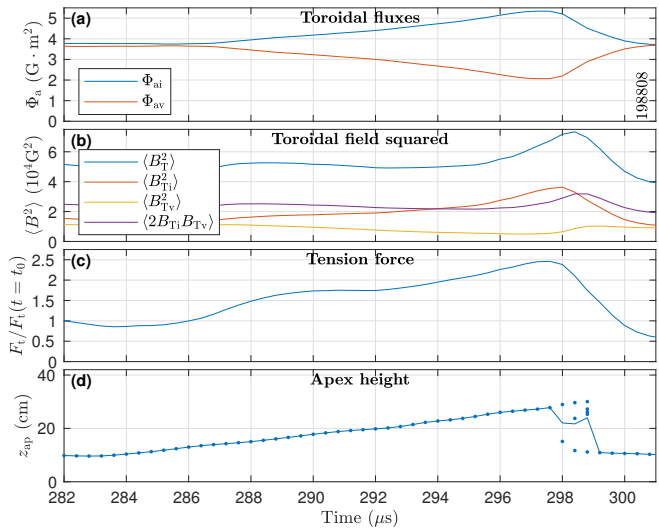}
    {\phantomsubcaption\label{fig:flux-indexed-a}}
    {\phantomsubcaption\label{fig:flux-indexed-b}}
    {\phantomsubcaption\label{fig:flux-indexed-c}}
    {\phantomsubcaption\label{fig:flux-indexed-d}}
    \caption{Various rope parameters related to flux and field profile during a sample failed torus event. Here the rope has been defined by a constant total toroidal flux, $\fluxT{a}$ rather than by the toroidal current. The tension force increases by a factor of more than two, consistent with experimental measurements of \Ft{}. \textbf{(a):} The flux due to the vacuum and internal fields. \textbf{(b):} The average value of the square of the total toroidal field compared with the vacuum and internal values. \textbf{(c):} The tension force derived from $\langle\BT^2\rangle$ normalized to the value at a reference time before the event. \textbf{(d):} The apex height for reference. During the failed torus event, the apex height is briefly multi-valued.}
    \label{fig:flux-indexed}
\end{figure}

To that end, we have analyzed the changes in $\langle \BT^2\rangle$ and the predicted \Ft{} during an example failed torus event in \cref{fig:flux-indexed}. Rather than defining the boundary of the rope by a fixed fraction of the total current, we have held the total toroidal flux, \fluxT{a}, constant in time. This means that the new field lines that are added to the rope are not being considered in this analysis, but rather a consistent flux surface is being considered. As the rope rises, \fluxT{a,v} decreases and \fluxT{a,i} increases as expected. We also see that $\langle \BT^2\rangle$ increases sharply during the collapse. This causes \Ft{} (defined here via \cref{eqn:FTpara}) to increase dramatically, which in turn causes the rope to collapse and the eruption to fail. By comparing $\langle\BT[i]^2\rangle$ and $\langle\BT[v]^2\rangle$ separately in \cref{fig:flux-indexed-c}, we see that this increase is largely driven by the change of the internal profile rather than the decrease in \Bg. The value of \Ft{} increases by a factor of more than two from the initial state to the peak during the collapse. This is consistent with experimental measurements where \Ft{} is calculated by integrating the $\bb{J}\btimes\bb{B}$ forces rather than using \cref{eqn:FTpara} \citec{myers2015}.

\subsection{Effects of the kink instability on failed torus events}
\label{sec:kink-on-FT}

By analyzing the change in the tension force, \Ft, in a rising MFR while holding the total toroidal flux at the apex, \fluxT{a}, constant, we have shown that failed torus effects can be explained entirely by ideal MHD without the need for adding flux in a non-ideal layer. We are left with the question as to why these effects are absent in the eruptive regime (the lower left quadrant of \cref{fig:ns_vs_q}) when the safety factor, \qa{}, is small. When a rope is kink-unstable, its axis is able to tilt in the $x$-$y$-plane, mixing the roles of the strapping and guide fields. In particular, if a rope's axis is tilted an angle of $\theta$, its effective strapping and guide fields would be
\begin{subequations}
\begin{align}
    B_\ms{s,eff} &= B_y\cos\theta -B_x\sin\theta ,\\
    B_\ms{g,eff} &= B_y\sin\theta +B_x\cos\theta,
\end{align}
\end{subequations}
where $B_y$ and $B_x$ are, respectively, the strapping and guide fields when there is no tilt. The change in a rope's guide field becomes conflated with the change in strapping field, and so the rope is able to erupt by slipping through the field lines in the arcade above it.

If instead a rope is kink-stable and cannot twist significantly, it must rise up through the arcade without tilting and experience a changing guide field based on the decay index of the guide field, $n_\ms{g}$. In order to begin the rise however, it must be initially unstable to the torus instability with $\ns>\ncr$ (limiting us to the upper right quadrant of \cref{fig:ns_vs_q}). Then when the guide field decays, the profile can change, enhancing \Ft. However, this assumes that \Bg{} decays fast enough for this effect to happen. For example, if $n_\ms{g}=0$, then \Bg{} would be a constant and we would expect no \Ft{} enhancement from these effects. This suggests that there may be a critical value of $n_\ms{g}$ where this effect could happen, i.e. it occurs when $n_\ms{g}>n_\ms{g,cr}$. However, due to experimental constraints, we were not able to probe to sufficiently small values of $n_\ms{g}$ to see these effects. Furthermore, on the Sun, \Bg{} is generated by external currents below the photosphere. Therefore small values of $n_\ms{g}$ may not be realizable on the Sun either.

Additionally, the movement of the apex may not be small when considering the onset of the failed torus, so rather than using the local derivative (or equivalently, $n_\ms{g}$), a finite difference should be considered instead. We empirically observe that the failed torus occurs when the change in guide field is around half of its initial value, i.e. $|\Delta \Bg|/\Bg[0]\sim 1/2$. For a given configuration with decaying \Bg{} this will occur at some finite displacement, $\Delta\zapex$, away from its initial apex height. The failed torus will be able to disrupt the eruption if $\Delta\zapex$ is small enough such that some other non-ideal effect is not able to disrupt this mechanism. For example, once a rope rises far enough, tether cutting reconnection can occur below the rope, disrupting the initial topology. This reasoning can lead to estimates for $n_\ms{g,cr}$. However, due to the complexity of the fields and the onset of reconnection, and analytical solution is intractable and it warrants future numerical study.

\subsection{Application to solar conditions}

The toroidal flux increase seen in our experiments in \cref{fig:apex-flux-during-event} may not be possible in solar flux ropes. In our experiments, a non-ideal layer above the electrodes allows for field-line bending to increase the total flux in a rope. However, using classical values for resistivity and viscosity in the photosphere, yields layers that are far too small to contribute a significant amount of flux (see \cref{tab:non-ideal-layers}). Despite the lack of flux increase, the results of \cref{sec:flux-indexed-values} and \cref{fig:flux-indexed} show that redistribution of \Bti[] can account for a significant increase of \Ft{} without any increase of \fluxT{a}. Since \Ft{} depends on $\langle\BT^2\rangle$ rather than the total flux, \Ft{} can be increased by making \BT{} less uniform, i.e. a \BT{} profile that is either more peaked or more hollow will increase $\langle\BT^2\rangle$.

We have discussed the differing roles of \ns{} and $n_\ms{g}$ in the onset of the torus instability and subsequent failed torus events and how these roles get mixed when kinking occurs at the apex. \ns{} is important for driving the torus instability while $n_\ms{g}$ can cause the necessary changes to create a failed torus and end the eruption. However, this may prove challenging for future predictions of solar events because observations are often unable to determine the direction of measured magnetic fields and therefore cannot independently measure \ns{} and $n_\ms{g}$ \citec{liu2008}.

\section{Discussion and conclusions}

The prediction of space weather requires a detailed understanding of the ideal MHD behavior of arched, line-tied, magnetic flux ropes. Models that are widely used to describe the behavior of toroidally symmetric systems such as tokamaks break down when extended to the line-tied system of an MFR. Laboratory experiments have been carried out to investigate the behavior of ideal MHD instabilities, particularly the torus instability, in MFRs.

In addition to the standard torus instability occurring at small values of \qa{}, a class of ``failed torus'' MFRs that were torus-unstable but kink-stable where observed failing to erupt \citec{myers2015}. These ropes often begin to rise due to the torus instability, but then the current profile becomes hollow, the toroidal field (and therefore the tension force) is increased and the eruption ultimately fails. We have investigated the energy inventory before and after these events and found that ideal MHD can explain the energy change to within 6\%, well within experimental error. In reversed field pinches, Taylor relaxation is often considered as the mechanism through which the current profile can change, resulting in a flat twist density profile, \mbox{$\alpha=\bb{J}\bcdot\bb{B}/(\mu_0B^2)$} \citec{taylor1974}. However, Taylor relaxation only conserves the total helicity and not the helicity of each flux surface. This means that internal reconnection (and therefore non-ideal MHD effects) is required for the plasma to reach its lowest energy state. This means that Taylor relaxation cannot be the cause of the current profile change in failed torus ropes.

By measuring the current distribution at the foot points during failed torus events, we have confirmed that the current redistribution is a global phenomenon and not localized to the apex where previous measurements were focused. We have also seen that, in our experiments, the total toroidal flux, \fluxT{a}, at the apex of a failed torus rope increases as it rises during an event. We have developed a model for how this can occur without violating the ideal MHD assumptions in the bulk of the rope. As a rope rises, the external guide field, \Bg, at the apex decreases. This causes the poloidal current, \JP, to increase away from the center. The resulting currents interacting with the frozen-in condition at the conducting electrodes, causes flux to be pulled into the rope thorough a thin non-ideal layer above the electrodes. The current is able to cross field lines in this layer and therefore new field lines are added to the rope, increasing \fluxT{a}. This flux increase then causes the tension force, \Ft, to increase, ultimately causing the eruption to fail. The expected flux changes during these events have been confirmed by measurements of flux loops placed below the electrodes. 

Additionally, changes in the toroidal field profile have been measured during failed torus events in our experiments. The changes in $\langle\BT^2\rangle$ when holding \fluxT{a} constant have been shown to be significant enough to cause \Ft{} to increase by a factor of more than two. Due to the thin non-ideal layers on the Sun, this redistribution mechanism is expected to be far more significant for solar flux ropes than any \fluxT{a} increase. This mechanism for failed eruptions could be used to predict failed CME events on the Sun.

\begin{acknowledgements}
    The authors would like to thank P. Sloboda and A. Jones for their help with the experimental setup. This research is supported by Department of Energy contract numbers DE-SC0019049 and DE-AC02-09CH11466 and NASA grant number 80HQTR17T0005.
\end{acknowledgements}

\section*{Data availability}
Data underlying the results presented in this paper will be made available via \href{https://dataspace.princeton.edu/handle/88435/dsp01pz50gz45g}{Princeton University's data repository} upon publication.

\appendix

\section{Conservation of energy in a flux rope under ideal MHD}
\label{sec:E-conservation}

We would like to analyze the energy inventory of an MFR before and after an eruptive or failed torus event. We will consider the energy equation using the assumptions of ideal MHD and then will compare to our experimental results in \cref{sec:LP-E-breakdown}.\footnote{The ideal MHD equations presented in this section neglect thermal conduction by assuming an adiabatic equation of state. However, the adiabatic assumption is confirmed to be valid in our experiments during the timescales discussed in \cref{sec:LP-E-breakdown}.} The agreement (or disagreement) will help validate (or invalidate) our assumptions. A substantial difference in the energy between the experiment and theory would indicate the need for non-ideal processes such as reconnection to explain the events.

The conservative form of the ideal MHD energy equation is given by \citec{freidbergBook2014}
\begin{equation}\label{eqn:Econ}
    \pD{t}{w}+\grad\bcdot\bb{s}=0,
\end{equation}
where $w$ and \bb{s} are given by
\begin{subequations}\label{eqn:Econ_w_s_def}
\begin{align}
    w&= \frac{1}{2}\rho v^2 +\frac{P}{\gamma-1} + \frac{B^2}{2\mu_0} ,\\
    \bb{s} &= \left(\frac{1}{2}\rho v^2 + \frac{P}{\gamma-1}\right)\bb{v} + P\bb{v} + \frac{1}{\mu_0}\bb{E}\btimes\bb{B},
\end{align}
\end{subequations}
where $\rho$ is the plasma mass density, $P$ is the plasma pressure, \bb{E} is the electric field, $\gamma$ is the adiabatic index, and \bb{v} is the plasma velocity. The terms in $w$ represent the density of the kinetic energy, the internal energy, and the magnetic energy, respectively. Whereas the terms of \bb{s} represent the flux of energy density (both kinetic and internal), the compressive work done on the plasma, and the Poynting vector, respectively. We will be assuming ideal MHD holds and so will use the ideal form of Ohm's Law,
\begin{equation}\label{eqn:Ohms-law}
    \bb{E} + \bb{v}\btimes\bb{B} = 0,
\end{equation}
i.e. we will assume that there is no resistivity.

The total energy in the system is given by
\begin{equation}
    W=\int_V w \rmd^3r,
\end{equation}
where the integral is carried out over the entire plasma volume, $V$. We then take a time-derivative of this equation, 
\begin{equation}\label{eqn:Econ_dWdt}
    \D{t}{W} = \int_V\pD{t}{w}\rmd^3r + \int_S w\bb{v}\bcdot\bb{\rmd S} ,
\end{equation}
where $S$ is the surface of the plasma volume. The surface, $S$, moves with the plasma velocity, \bb{v}, and so the rate of change of the volume is given by
\begin{equation}
    \D{t}{V} = \int_S\bb{v}\bcdot\bb{\rmd S}.
\end{equation}

We can then substitute \cref{eqn:Econ,eqn:Econ_w_s_def} into \cref{eqn:Econ_dWdt} and use \cref{eqn:Ohms-law} to get (See \citet{alt2022thesis} for a more detailed derivation.)
\begin{equation} \label{eqn:E_conservation} 
    \D{t}{W} = -\int_S \left(\bb{s}-w\bb{v}\right)\bcdot\bb{\rmd S} = -\int_S \left(P+\frac{B^2}{2\mu_0}\right)\bb{v}\bcdot\bb{\rmd S} .
\end{equation}
This equation implies that the change in total energy in the system is equivalent to the work done by compression of both the thermal and magnetic pressures.

\bibliography{MFR-bibliography}
\end{document}